\documentclass[aps,superscriptaddress,preprint,preprintnumbers,amsmath,amssymb,onecolumn]{revtex4-1}

\usepackage[utf8]{inputenc}

\usepackage{graphicx}
\DeclareGraphicsExtensions{.pdf,.jpg,.png,.tif}

\usepackage{dcolumn}
\usepackage{bm}
\usepackage{hyperref}
\usepackage{mathrsfs}
\usepackage{epsfig}
\usepackage{epstopdf}
\usepackage{multirow}
\usepackage{amssymb}
\usepackage{amsmath}

\usepackage{color}
\newcommand\p{\ensuremath{\partial}}
\newcommand\red{\color{red}}

\newcommand\bx{\bm{x}}
\newcommand\bxo{\bm{x_0}}

\begin{document}
%
%
\title{Deformation and stability of a viscous electrolyte drop in a uniform electric field}

\author{Qiming Wang}
\affiliation{Department of Mathematical Sciences and Center for Applied Mathematics and Statistics, New Jersey Institute of Technology, Newark, NJ 07102, USA}

\author{Manman Ma}\email[Corresponding author: ]{mamm@tongji.edu.cn}
\affiliation{School of Mathematical Sciences, Tongji University, Shanghai 200092, China}

\author{Michael Siegel}
\affiliation{Department of Mathematical Sciences and Center for Applied Mathematics and Statistics, New Jersey Institute of Technology, Newark, NJ 07102, USA}


\begin{abstract}
We study the deformation and breakup of an axisymmetric electrolyte drop which is freely suspended in an infinite dielectric medium and subjected to an imposed electric field. The electric potential in the drop phase is assumed small, so that its governing equation is approximated by  a linearized Poisson-Boltzmann or modified Helmholtz equation (the  Debye-H\"{u}ckel regime). An accurate and efficient boundary integral method is developed  to solve the low-Reynolds-number flow problem for the time-dependent drop deformation, in the case of arbitrary Debye layer thickness. Extensive numerical results are presented  for the case when the viscosity of the drop and surrounding medium are comparable.
Qualitative similarities are found between the evolution of a drop with a thick Debye layer (characterized by the parameter $\chi\ll 1$, which is an inverse dimensionless Debye layer thickness) and a perfect dielectric drop in an insulating medium. In this limit, a  highly elongated steady state is obtained for sufficiently large imposed electric field, and the field inside the drop is found to be  well approximated using slender body theory.
In the opposite limit $\chi\gg 1$, when the Debye layer is thin, the drop behaves as a highly conducting drop, even for moderate permittivity ratio $Q=\epsilon_1/\epsilon_2$, where $\epsilon_1, \epsilon_2$ is the dielectric permittivity of drop interior and exterior, respectively. 
For parameter values at which steady solutions no longer exist, we find three distinct types of unsteady solution or breakup modes.  These are termed conical end formation, end splashing, and open end stretching. The second breakup mode, end splashing,  resembles  the breakup solution presented in a recent paper [R. B. Karyappa {\it et al}., J. Fluid Mech. {\bf 754}, 550-589 (2014)]. 
We compute a phase diagram which illustrates the regions in parameter space in which the different breakup modes occur.
\end{abstract}

\maketitle

\section{\label{section:Introduction}Introduction}
The behavior of a viscous liquid drop immersed in a viscous surrounding fluid and acted on by an imposed electric field is a classical problem which has been extensively studied  for over one hundred years. It is known that a mismatch in electrical properties 
between the fluids results in a jump in electric stress at  the drop interface. In the case of a drop subjected to a uniform far-field electric field, non-uniform tractions at the drop surface lead to  deformation of the interface and, for a sufficiently large imposed field, breakup of the drop. This problem arises in a number of important applications, including electrosprays, electrohydrodynamic atomization, breakup of droplets in thunderstorms, microfluidic processes, and others. A thorough review of the topic can be found in~\cite{Melcher81,MelcherTaylor,LeeReview}.

For either the case of a perfect dielectric or a perfectly conducting drop in an insulating medium
the electric field  modifies the normal stress at the interface  but does not affect the  tangential stress. The normal interfacial electric stress is balanced by surface tension,  but the lack of a tangential electric stress to balance viscous stresses  means that  there can be no fluid flow when the drop reaches a steady-state shape. This leads to  prolate steady-state drop shapes, as shown in a number of early studies~\cite{Zeleny,Macky,WilsonTaylor,Allan62,Taylor64,Torza71}.  
A theoretical prediction of the steady-state solution branch based on a spheroidal approximation of the equilibrium drop shape is given in \cite{Taylor64,Garton64}.  Above a critical permittivity ratio  $Q^c \simeq 20.8$, the theoretical steady solution branch  forms an 'S' shape and  is no longer single valued.   Time-dependent boundary integral computations \cite{Sherwood88}  of dielectric or non-conducting drops for $Q$  greater than  this critical value, starting from a spherical shape, converge to   steady state profiles on the solution branch  for  
sufficiently small electric field strength as measured by the electric capillary number  $E_b=\epsilon_2\mathscr{E}^2R/\gamma$, where $\mathscr{E}$ is the applied field strength, $R$ is the undeformed drop radius, and $\gamma$ is the surface tension.   However, beyond a  critical electric field strength $E_{b}^c$ associated with  the first turning point on the solution branch, the evolving drop forms a pointed tip and  the numerics break down before reaching the steady-state shape.   
This gives, in effect,  a  critical permittivity ratio $Q^c$ and  electric field strength $E_{b}^c$ above  which steady drop shapes are no longer observed.  We note that steady-state shapes on the upper branch could be obtained in the numerics if the initial drop shape was sufficiently close to equilibrium \cite{Sherwood88}, or by using other models \cite{Pillai15}.  When $Q$ is less than the critical value, the boundary integral computations converge to the steady state for any electric field strength.

The pointed drop shape that occurs above a critical permittivity ratio $Q$ and electric field strength was first investigated by G. I. Taylor for conducting water droplets in \cite{Taylor64} and is known as a Taylor cone.  Later
theoretical work includes predictions  of  the conical-end angles
of a Taylor cone \cite{SLB99, BFKV2006,FKV2008,GMT2017}.
It is worth noting that Taylor's analysis~\cite{Taylor64} is in fact based on a local solution that assumes a steady or equilibrium cone shape 
while in experiments the dynamics is often observed to be unsteady, with a thin, charged fluid jet emitted from the end of the conical or pointed tip in a process known as tipstreaming \cite{Basaran08NP}.
A recent review on the topic of Taylor cones in two-phase flow can be found in~\cite{Mora07}. 

In addition to the prolate drop profiles mentioned above,
experiments~\cite{Allan62} in weakly conducting fluids  show the presence of  oblate drop shapes and non-zero fluid velocity even after a drop has reached a steady-state shape. These characteristics are not
captured by the  perfect dielectric and perfect conductor  models. 
To  explain these features, Taylor  proposed the so-called leaky dielectric \cite{Taylor66} or Taylor-Melcher (TM) model \cite{MelcherTaylor} for weakly conducting fluids, which has been widely and successfully applied to many problems in electrohydrodynamics (see,  e.g., \cite{CrMat2005,Tseluiko08PoF, Mestel96,Hohman2001a,Wang2012,Ding2013,WP2016} in addition to the references below). The leaky dielectric model   allows charge to accumulate at a fluid-fluid interface, 
and tangential electric stresses generated by this surface charge along with charge convection are found to be important for  predicting oblate  steady-state drop shapes, steady fluid motion, and unsteady breakup in isolated drops acted on by an electric field \cite{collins2013, Sengupta2017}. The  model has been used extensively in numerical studies of the steady deformation and unsteady breakup of an isolated drop in an imposed electric field, see  e.g., \cite{collins2013, Feng1996, Hua2008, LacHomsy07, Sengupta2017, Sherwood88}. 
The results of these studies are in qualitative agreement with experiments involving weak electrolytes, although quantitative agreement is sometimes lacking. 
In particular, the numerical  simulations capture the  two main drop breakup modes observed in experiments \cite{Allan62, Torza71, HY2000a,HY2000b}. These are  (i) end pinching, in which the drop elongates and forms fluid blobs at its ends that eventually pinch off,
and (ii) tipstreaming, i.e., the formation of a Taylor cone, followed by emission of a thin charged fluid jet or a series of small droplets from the pointed tip.
Tipstreaming from a  viscous drop or fluid layer  in an imposed electric field  has  been  studied by  finite element numerical simulations of the leaky dielectric model  \cite{Basaran08NP, collins2013}. 
 In particular, these studies demonstrate the importance of charge convection and tangential electric  stress at the interface in the phenomena of  fluid ejection from Taylor cones.  
However, various experiments also report  discrepancies between the TM model  and experimental results (see for example~\citet{VS1992} and \citet{HY2000a}), which suggest the necessity of further modeling work.
 A thorough review on the TM model can be found in~\citet{MelcherTaylor,Saville1997}, and~\citet{Vlahovska16PRF}.

In recent years there has been renewed interest in  more detailed electrokinetic models for systems of drops and bubbles which  incorporate equations governing the dynamics of bulk ionic charge.  Theoretical studies have often  focussed on the asymptotic analysis  
 in the  thin Debye layer limit of the electrokinetic equations for various physical problems \cite{BaygentsSaville89, MY2017, Pascall2011, SY2015}. 
There has been much less work on the direct numerical simulation of the full electrokinetic equations.  Berry et al. \cite{Berry2013} developed a combined level-set/volume-of-fluid method to simulate the Poisson-Nernst-Planck electrokinetic model for liquid-liquid interfaces.  
 Pillai ~\cite{Pillai15,Pillai16} employed the method of \cite{Berry2013} to study the deformation and breakup of an isolated electrolyte drop suspended in an insulating phase and acted on by an imposed electric field.  Their results include computations of tip streaming drops. Related numerical work includes \cite{Davidson2014, Lopez2015, pillai2017}.  A strong motivation for  theoretical and numerical analysis of the electrokinetic models is their significance in microfluidic devices, for which  electrokinetic techniques have been among the most important methods for the manipulation of drops and bubbles \cite{Squires2005}. 

In this  paper, we modify the traditional model for a perfect dielectric drop in an insulating medium by assuming that the fluid inside the drop is an electrolyte, while keeping the exterior fluid as nonionic. We avoid the difficulty of solving the full nonlinear problem for the ion dynamics,  as in~\cite{Pillai15, Pillai16}, by employing the Debye-H\"{u}ckel approximation for the electric potential inside the drop, which results in  a linearized Poisson-Boltzmann or  modified Helmholtz equation.  Our model is therefore less general than the full electrokinetic model of \cite{Pillai15, Pillai16}, but has the advantage that the governing equations allow a Green's function formulation, which 
enables the development and application of a  highly accurate boundary integral numerical method. This surface based numerical method 
can effectively compute 
  for much thinner Debye layers (i.e., for $\chi \gg 1$)  than is possible for the full electrokinetic model.  
 A similar model was previously presented in~\citet{HLK10}, but there the focus is on analytical theory for small drop deformation. We go beyond this and carry out a more comprehensive numerical investigation.

Numerical computations based on a boundary integral formulation for  the problem of freely suspended drops in an electric field have been popular due to their high accuracy and relative simplicity ~\cite{Miksis1981,Sherwood88,BFKV2006, DubashMestel07, LacHomsy07,DT2012,KDT2014,Das2017,Sengupta2017}. When the electric potential is governed by  Laplace's equation, there is an analytical expression for the axisymmetric version of the Green's function, i.e., the azimuthal part of the surface integral can be done analytically.  This reduces the dimension of the  boundary integrals  and leads to a significant reduction in computational cost. However, this is not the case for the modified Helmholtz equation that arises here from linearizing  the Poisson-Boltzmann equation.   Although accurate numerical schemes to solve boundary integral formulations of Laplace and Helmholtz equations in axisymmetric geometries have been developed (see~\cite{Martinsson2010} and references therein), one of the main contributions in this paper is to develop a scheme to accurately and efficiently compute the Green's function for the modified Helmholtz equation and apply it to a moving boundary problem.
In this way, we extend previous studies to assess the effect of ions on drop deformation in the case of 
arbitrary Debye-layer thickness.  For numerical efficiency, our results are specialized to the case when the viscosity of the drop and surrounding medium are comparable.

Along with the numerical simulations, we also carry out a slender-body analysis (in the case of highly elongated drops) starting from the boundary-integral equations to approximate the electric field inside the drop. A correction term to the results of~\cite{SLB99} that takes into account the presence of ions  is derived, and the result is shown to agree reasonably well with simulations of the  full problem.

The analysis and numerics are used to find steady state solution branches and unsteady breakup modes over a wide range of parameter values. These solutions are found to be in good agreement with previous computations \cite{DubashMestel07, FKV2008, Sherwood88}  in the limiting cases of a perfect conductor or perfect dielectric, but exhibit some differences  
when compared to simulations based on the full electrokinetic model in \cite{Pillai15}.   
For parameter values at which steady solutions no longer exist, we find three distinct types of unsteady solution or breakup modes, which are termed conical end formation, end splashing, and open end stretching (see Section \ref{sec:breakup} for examples of these breakup modes). 
Similar breakup phenomena have been previously reported in 
simulations of perfect conductor, perfect dielectric, and leaky dielectric models ~\cite{FKV2008, KDT2014, Sherwood88, Sengupta2017}.  However,  there are some important differences in the results here. 

To maintain simplicity, we do not allow for charge to accumulate at the interface.  Thus, like the perfect dielectric and perfect conductor models,
there is no tangential electric stress at the interface (see Section \ref{sec:equations} for a detailed discussion). 
The general expectation   is that drop breakup by tipstreaming or pinch-off does not occur in the absence of such  tangential interface stresses \cite{collins2013}. 
Although our model does not capture  the tip streaming of thin charged jets from a Taylor cone, as in  \cite{collins2013, Pillai15}, a main result of this paper  is the computation of breakup modes without tangential interface stress, as  shown in Section \ref{sec:breakup}.  One of these breakup modes, end splashing, involves the formation of a Taylor-cone-like structure, followed by the emission or discharge of a thin axisymetric sheet of fluid which subsequently can pinch off.   This breakup mode is not seen in computations of the leaky dielectric model, but is very similar to that observed in experiments and simulations  of an electrokinetic model in  \cite{Mohamed2016}, 
based on  a pendant drop geometry. A similar breakup mode is  observed in experiments on high conductivity drops and simulations of a perfect conductor model at small viscosity ratio ($Q \rightarrow \infty$ and $\lambda \lesssim 0.05$) in \cite{KDT2014}. Here, however, we find  that  end-splashing also  occurs at a finite conductivity ratio $Q$ 
and for unit viscosity ratio $\lambda=1$, suggesting  that electrokinetic effects  can promote this type of breakup.  
Preliminary  computations  using our model at a small viscosity ratio, to be reported elsewhere, also exhibit the end-splashing breakup mode.

The model presented here is formally valid for small potential, or more precisely, when $e\phi / (k_BT)\ll1$. However, we will sometimes also apply it to cases with large deformation and potentials (e.g., Figure~\ref{tipb}) which may be beyond its range of formal validity.  Nevertheless,  qualitative similarities  between solutions to our model and behavior  observed in the experiments of \cite{Mohamed2016, KDT2014}, even at relatively large potentials,  are encouraging.   A discussion of the Debye-H\"{u}ckel theory and its limitations is given in~\cite{Wright2007}. 

The paper is organized as follows. We begin in Section~\ref{sec:equations} with a complete description of the equations governing the electric field, viscous flow and the boundary conditions. In Section~\ref{sec:bie}, the problem is reformulated as a system of boundary  integral equations and the numerical method is introduced. Numerical results are presented in Section~\ref{sec:results}. We summarize the effect of ions on the drop's steady deformation and unsteady breakup behavior. 
Closing remarks are provided in Sec.~\ref{sec:concl}. In Appendix~\ref{Ggrad}, we present the formulations of the Green's function and its derivatives for the modified Helmholtz equation, as well as results demonstrating the accuracy of our numerical scheme in computing the Green's functions. In Appendix~\ref{smd}, we present a brief derivation of the small deformation theory for our problem.  This is used to compare with and partially validate our numerical results.   Finally, details of the  slender-body analysis are presented in Appendix~\ref{slendapp}.

\section{Mathematical formulation}\label{sec:equations}
\subsection{Electrokinetic equations}
\begin{figure}[!ht]
\includegraphics[scale=0.5]{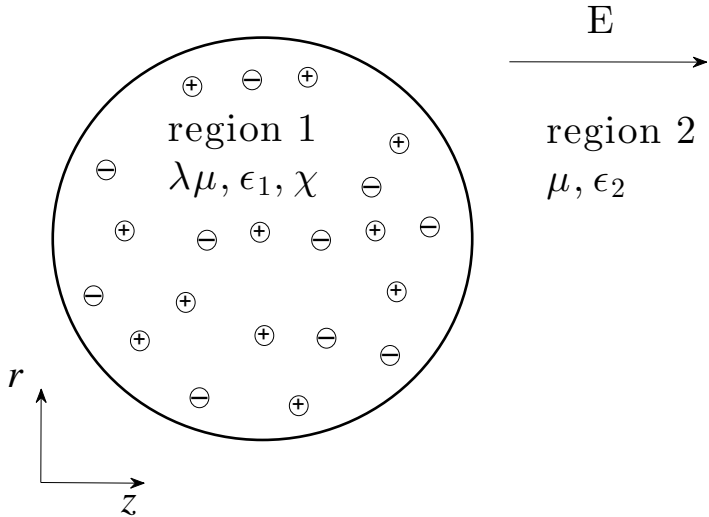}
\caption{An electrolyte fluid drop with viscosity $\lambda \mu$ is surrounded by a nonionic fluid with viscosity $\mu$. A constant electric field, directed along the $z$-axis,  is imposed in the far field.}\label{domain}
\end{figure}

We consider the dynamics of an electrolyte fluid drop with viscosity $\lambda\mu$ (region 1) immersed in  a dielectric (nonionic) medium with viscosity $\mu$ (region 2),  as shown in Figure~\ref{domain}. Cylindrical polar coordinates $\bm{x}=r\bm{e}_r+z\bm{e}_z$ are used with the $z$-axis  alligned with the drop's axis of symmetry. 
The surrounding medium is considered as a perfect dielectric and the electric potential $\phi_2$ satisfies Laplace equation with a far field condition $\phi_2\rightarrow -\mathscr{E}z$ due to the applied electric field ${\bf E} = \mathscr{E}\bm{e}_z$.

In the drop phase, the electric potential is governed by Poisson's equation,
\begin{equation}
-\epsilon_1\nabla^2\phi_1= \rho = \sum_{i=1}^N ez_ic_i,\label{poisson}
\end{equation}
with $\epsilon_1$ the dielectric constant,  $\rho$ the bulk volume charge density, $z_i$ the valence of species $i$ and $e$ the elementary charge. We assume that the ions are in thermo-equilibrium and that their concentration  follows a  Boltzmann-distribution~\cite{HinchSherwood83,Conroy10JFM,Mao2014JFM},
\begin{equation} \label{thermoeq}
 c_i = c^0_ie^{-z_ie(\phi_1-\phi_0)/(k_BT)},
\end{equation}
 where $\phi_0$ is a reference potential, 
 which is set to zero without loss of generality, $k_B$ is the Boltzmann constant, and $T$ is the absolute temperature.  We consider a symmetrical electrolyte with $z_1=-z_2=z$ and  $c^0_1=c^0_2=c^0$, which leads to an odd symmetry of the potential $\phi_1$ with respect to the  plane $z=0$,  
where $\phi_1\equiv 0$.  We take $c^0_i$ to be the constant bulk concentration of ion species $i$ far from the interface when this bulk state exists (e.g., for a thin Debye layer), and otherwise take it to be the ion concentration on the plane $z=0$.  The neutral bulk condition $\sum_{i=1}^2 z_ic_i^0=0$ is assumed to hold.  We introduce
\begin{equation}
\beta = \frac{\mathscr{E}eR}{k_B T}
\end{equation}
which measures the ratio of the imposed field potential over the thermal potential.
Using $R$, $\mathscr{E}R$ and  $c^*$ as the characteristic length, electric potential and ion concentration in equation (\ref{poisson}), where $R$ is the unperturbed drop radius, $\mathscr{E}$ is the uniform electric field at infinity, and $c^*=\int_\Omega c_i  \ dV/\int_\Omega   dV$ is the average concentration of  ion species $i$ in drop region $\Omega$ (which is the same for $i=1,2$),  we obtain the Poisson-Boltzmann equation in dimensionless form

\begin{equation}
 -\nabla^2\phi_1= \frac{eRc^*}{\epsilon_1 \mathscr{E}}\sum_{i=1}^2 z_i c^0_ie^{-\beta z_i\phi_1}\label{poisson1}.
\end{equation}
In the case of  small applied  (drop phase) potential $|\beta\phi_1|\ll 1$, the ion concentration is approximated as $c_i\approx c_i^0(1-\beta z_i\phi_1)$ and ionic mass conservation in the drop combined with symmetry of the potential $\phi_1$  gives $c_i^0=1$. Thus, Eq. \eqref{poisson1} simplifies at leading order  to the linearized Poisson-Boltzmann or modified Helmholtz equation,
\begin{equation}
\nabla^2\phi_1= \chi^2\phi_1,\label{poisson2}
\end{equation}
where 
\begin{equation}
\chi^2  
= \frac{2z^2e^2R^2c^*}{\epsilon_1 k_B T}.
\end{equation}
This linearized equation, also known as Debye-H\"{u}ckel approximation, is widely used in various problems involving electrolyte solutions~\citep{White78, HinchSherwood83, BaygentsSaville91a, SS1994PoF, HLK10, Conroy10JFM, Mao2014JFM}.  An advantage of (\ref{poisson2}) is that there is a Green's function representation of the solution, which is evaluated using a highly accurate boundary integral numerical method. Since drop breakup can require large applied voltage, we will sometimes  apply the approximation (\ref{poisson2})  in situations in which it is not formally valid.  

At the drop interface, we have the boundary conditions 
\begin{align}
\phi_1=\phi_2,\quad Q\phi_{1n}=\phi_{2n},\label{eBC}
\end{align}
where $Q=\epsilon_1/\epsilon_2$. This specifies that  no  ionic charge accumulates  at the interface, a condition used in the electrolyte-drop model of~\cite{HLK10}.

\subsection{Fluid motion and stress boundary conditions}
The fluid motion is approximated by Stokes equations, which are nondimensionalized using the spherical drop radius $R$ for lengths, $\gamma/R$ for pressure, {$\mathscr{E}R$ for potential}, and  $\gamma/\mu$ for velocities, with $\gamma$ the surface tension. {After replacing $\nabla^2\phi_1$ by $\chi^2\phi_1$ from \eqref{poisson2}, we have}  
\begin{align}
& -\nabla p_1 + \alpha \phi_1 \nabla\phi_1 +\lambda\nabla^2 \bm{u}_1 = 0,\quad \nabla \cdot\bm{u}_1=0,\label{stokes1}\\
 &-\nabla p_2 + \nabla^2 \bm{u}_2 = 0,\quad \nabla \cdot\bm{u}_2=0.
\end{align}
In the above, 
\begin{equation} \label{alpha_Eb}
\alpha=\chi^2E_bQ  ~~\mbox{with} ~~ E_b=\epsilon_2\mathscr{E}^2R/\gamma,
\end{equation}
where the latter quantity is an {electric capillary number} which  measures the ratio of Maxwell  or electric stress  to capillary pressure.  The stress balance on the interface is written as
\begin{equation}\label{Tn}
[\bm{\mathcal{T}}\cdot \bm{n}]_2^1=[\bm{\mathcal{\sigma}}\cdot \bm{n}]_2^1-\triangle \bm{f}^e=-\kappa\bm{n},
\end{equation}
where $\bm{\sigma}=-p  \bm{I} + 2 \lambda_i \bm{e}$ is  the hydrodynamic stress tensor with $\lambda_1=\lambda$ and $\lambda_2=1$, $\bm{e}$ is the symmetric part of the velocity gradient,
$\bm{n}$ is the outward unit normal,  and $\kappa$ is  the interface curvature, taken as positive for a convex surface.  Here $[ \cdot ]_2^1$ denotes the jump across the interface, 
with the convention that it is the limit as the interface is approached from  the interior domain (region 1) minus the limit from the exterior domain (region 2). 
The Maxwell stress, or electric contribution to the  stress balance, is directed normally to the interface and is given by
\begin{equation} \label{dfe}
\triangle \bm{f}^e=\frac{E_b(Q-1)}{2} \left(QE_{1n}^2+E_t^2 \right)\bm{n},
\end{equation}
where {$E_{1n}$} and $E_t$ are normal and tangential derivatives of electric field {in region 1} respectively (see  \citet{Miksis1981, Sherwood88, LacHomsy07}).  {The jump in tangential stress at the interface is 
{$ \epsilon_2E_t(E_{2n}-QE_{1n})$} 
\cite{Sherwood88}  and is zero in view of the boundary condition (\ref{eBC}) for a charge free surface. } Far from the drop, $\phi_2\rightarrow -z$ as $|\bm{x}|\rightarrow \infty$.

\section{Boundary integral method}\label{sec:bie}

\subsection{Integral equations}
We reformulate the electrostatic problem as a system of boundary integral equations using classical potential theory \cite{Kress}.  Denote the Green's function for the modified Helmholtz equation by $G^{\chi}$; expressions for this Green's function are presented in Appendix \ref{Ggrad}. The electric potentials $\phi_1$ and $\phi_2$ in regions 1 and 2 satisfy
\begin{align}
\frac{1}{2}\phi_1(\bxo)+\int_{S}\phi_1(\bx)\frac{\p G^{\chi}}{\p n_x}  (\bx,\bxo) \ dS(\bx) &= \int_{S}\frac{\p\phi_1}{\p n} (\bx) G^{\chi} (\bx,\bxo) \ dS(\bx),\label{pe1}\\
-\frac{1}{2}\left(\phi_2(\bm{x}_0)-\phi_{\infty} (\bxo)\right) &+\int_{S}\left(\phi_2 (\bx)-\phi_{\infty} (\bx)\right)\frac{\p G^0}{\p n_x} (\bx,\bxo)  \ dS (\bx),  \nonumber \\
&= \int_{S}\frac{\p \left(\phi_{2}  -\phi_{\infty} \right)}{\p n} (\bx)  G^0 (\bx,\bxo) \ dS(\bx),\label{pe2}
\end{align}
where $\phi_{\infty}=-z$ is the imposed far-field electric potential.  
We note that setting $\chi=0$ in (\ref{pe1}) recovers the case of a perfect dielectric drop,  in which electrolyte is not present in the interior (see \cite{Miksis1981}).

The standard boundary integral formulation of the  Stokes flow  problem for the fluid velocity is modified to include the electrostatic forcing. Starting from the Lorentz reciprocal relation   \cite{Poz1992}, we obtain  
\begin{align}
{\bm{u}_1(\bm{x}_0)}&= \frac{\alpha}{8\pi\lambda}\int_V \phi_1(\bx) \nabla\phi_1(\bx)   \cdot \bm{J}(\bx,\bxo) \ dV(\bx) + \frac{1}{8\pi\lambda}\int_{S}{\bm{n}(\bx) \cdot\bm{\mathcal{\sigma}}_1 (\bx) \cdot\bm{J}} (\bx,\bxo)  \ dS(\bx) \nonumber \\
&\qquad  \qquad+\frac{1}{8\pi}\int_{S} {\bm{n}(\bx) \cdot \bm{K} (\bx,\bxo) \cdot \bm{u}} (\bx)  \ dS(\bx),\label{bie_u}\\
{\bm{u}_2(\bm{x}_0)}&=  -\frac{1}{8\pi} \left(\int_{S}{\bm{n}(\bx) \cdot\bm{\mathcal{\sigma}}_2 (\bx) \cdot\bm{J}}(\bx,\bxo) \  dS(\bx)
+\int_{S} {\bm{n}(\bx) \cdot \bm{K}(\bx,\bxo) \cdot \bm{u}}(\bx) \ dS(\bx) \right),\label{bie_u1}
\end{align}
where $\bm{J}$ and $\bm{K}$ are the Stokeslet and Stresslet Green's functions for Stokes flow, {and $\bm{x}_0$ is located in regions 1 and 2 in \eqref{bie_u} and \eqref{bie_u1}, respectively}. The first term on the right hand side of (\ref{bie_u})  is transformed into a surface integral by using the divergence free property of the Stokeslet, namely,
\begin{equation}
\nabla\cdot \bm{J}=0,
\end{equation}
which leads to
\begin{align}
\int_V \phi_1(\bx) \nabla\phi_1(\bx)  \bm{J}(\bx,\bxo)  \ dV(\bx) &=\frac{1}{2}\int_V {\nabla_{\bx}} \cdot\left(\phi^2_1 (\bx)  \bm{J}(\bx,\bxo)\right) \ dV(\bx), \label{divphiJ}\\
&=\frac{1}{2}\int_S \phi_1^2 (\bx) \bm{J}(\bx, \bxo) \cdot\bm{n} (\bx)  \ dS(\bx), \label{phiJ}
\end{align}
{where \eqref{phiJ} follows \eqref{divphiJ} by the divergence theorem.}
As $\bm{x}_0$ approaches interface, the integral equations can be combined to one equation by using (\ref{Tn}),
\begin{multline}
{\bm{u}(\bm{x}_0)}=  -\frac{1}{4\pi(1+\lambda)}\int_{S}{\bm{J} (\bx,\bxo) \cdot \triangle \bm{F}^e} (\bx) \ dS(\bx) \\ -\frac{1-\lambda}{4\pi (1+\lambda)}\int_{S} {\bm{n}(\bx)  \cdot \bm{T}(\bx, \bxo)  \cdot \bm{u}} (\bx) \  dS(\bx), \label{bie_u2}
\end{multline}
where $\phi_s$ is the electric potential on  the interface $S$ and
\begin{equation}
\triangle \bm{F}^e=\left(\kappa - \frac{\alpha}{2}\phi^2_s\right)\bm{n}-\triangle \bm{f}^e,\label{dFe}
\end{equation}
where $\alpha$ is given in (\ref{alpha_Eb}) and $\triangle \bm{f}^e$ is given by (\ref{dfe}).
Similar integral equation formulations for a viscous drop in an  electric field have appeared in  \cite{Miksis1981,Sherwood88,LacHomsy07}, and \cite{Poz1992,Poz2002} provides  more details in the  derivation, as well as numerical implementations. Note that  the additional term $\alpha\phi^2_s$ in (\ref{dFe}) is a consequence of  the electric body force in the Stokes equation. This term can also be understood via a modified or effective pressure  $\hat{p}_1=p_1 - \frac{\alpha}{2}\phi_1^2$ in  (\ref{stokes1}),  which similarly implies (\ref{dFe}).
In the limit of a perfect conductor ($\chi\rightarrow \infty$), $\phi_1=0$, whereas in the limit of a perfect dielectric drop, $\chi\rightarrow 0$, hence $\alpha\rightarrow 0$. Therefore, this additional term vanishes in both extreme cases.

\subsection{Computation of Green's functions} \label{sec:Comp_Greens}
The axisymmetric version of the free space Green's function for Laplace's equation (defined as the azimuthal integral of the 3D Green's function)  can be expressed in closed form, see for example,~\cite{Poz2002},
\begin{align}
G^0(z,z_0,r,r_0)=\int_0^{2\pi}G^{3D} (z, z_0,r,r_0,\phi,\phi_0) d\varphi=\frac{K(k)}{\pi\sqrt{(z-z_0)^2+(r+r_0)^2}}, \label{axlg}
\end{align}
where $G^{3D} (z, z_0,r,r_0,\phi,\phi_0)=((z-z_0)^2+r^2+r_0^2-2 r r_0 \cos(\phi-\phi_0))^{-{1/2}}/(4\pi)$ is the Green's function for the 3D Laplace's equation in cylindrical coordinates,   $k^2=4rr_0/\left[(z-z_0)^2+(r+r_0)^2\right]$ and {$k\le 1$}. Here $K(k)$ is the complete elliptical integral function of first kind,
\begin{align}
K(k)=\int_{0}^{\pi/2}\frac{d\theta}{\sqrt{1-k^2\cos^2\theta}}.\label{kfun}
\end{align}
The axisymmetric Green's function for the  modified Helmholtz equation, however, does not have an analytical expression. Starting from the Green's function for the modified Helmholtz equation in cylindrical coordinates, we write the axisymmetric version as follows:
\begin{align}
G^{\chi}(z,z_0,r,r_0)&=\frac{1}{4\pi}\int_0^{2\pi}\frac{{\rm{exp}}\left(-\chi\left[(z-z_0)^2+r^2+r_0^2-2rr_0\cos u \right]^{1/2} \right)}{\left((z-z_0)^2+r^2+r_0^2-2rr_0\cos u \right)^{1/2}} du\nonumber\\
&=\frac{1}{4\pi}\int_0^{2\pi}\frac{{\rm{exp}}\left(-\chi\left[(z-z_0)^2+(r+r_0)^2-4rr_0\cos^2 (u/2) \right]^{1/2} \right)}{\left((z-z_0)^2+(r+r_0)^2-4rr_0\cos^2 (u/2) \right)^{1/2}} du\nonumber\\
&=\frac{k}{2\pi(rr_0)^{1/2}}\int_0^{\pi/2}\frac{{\rm{exp}}\left(-\Lambda\left[1-k^2\cos^2 \theta \right]^{1/2} \right)}{\left(1-k^2\cos^2 \theta \right)^{1/2}} d\theta,\label{axhmg}
\end{align}
where $\Lambda=2\chi (rr_0)^{1/2}/k$. When $|\Lambda|\ll 1$ or $|\Lambda|\gg 1$ in (\ref{axhmg}), the Green's function is expanded in an appropriate series for the numerical calculations in \cite{Priede2006}. In the present study, we focus on the direct evaluation of (\ref{axhmg}) by proper quadrature. 

Substitution of $t=\cos\theta$ into (\ref{axhmg}) results in 
\begin{align}
G^{\chi}(z,z_0,r,r_0)=\frac{k}{2\pi(rr_0)^{1/2}}\int_0^{1}\frac{{\rm{exp}}\left(-\Lambda\left[1-k^2t^2\right]^{1/2} \right)}{\left(1-k^2t^2\right)^{1/2}} \frac{dt}{(1-t^2)^{1/2}}.\label{axhmg1}
\end{align}
Gauss-Chebyshev quadrature would seem a natural choice to integrate 
(\ref{axhmg1}), treating   $(1-t^2)^{-1/2}$ as the weight function. However, our experience shows that Alpert quadrature~\citep{Alpert99} gives faster convergence and better performance. By recognizing $(1-t)^{-1/2}$ as a singular function inside the integrand and setting $t=1-x$, (\ref{axhmg1}) can be rewritten as
\begin{align}
G^{\chi}(z,z_0,r,r_0)=\frac{k}{2\pi(rr_0)^{1/2}}\int_0^{1}\frac{{\rm{exp}}\left(-\Lambda\left[1-k^2(1-x)^2\right]^{1/2} \right)}{\left(1-k^2(1-x)^2\right)^{1/2} (2-x)^{1/2} }x^{-1/2} dx.
\end{align}
Alpert quadrature uses a hybrid Gauss-trapezoidal quadrature rule for the integration $\int_0^1 f(x)dx$ where $f(x)= g(x)x^{-1/2}$ and $g(x)$ is regular. The quadrature follows the formula (see~\cite{Alpert99} for more details)
\begin{equation}
T^{jkab}_n(f) = h\left( \sum_{i=1}^{j}u_if(v_ih) + \sum_{i=0}^{m-1}f(ah+ih) + \sum_{i=1}^{k}w_if(1-x_ih)\right),
 \end{equation} 
where the nodes $ v_1, ..., v_j, x_1,..., x_k$ and weights $u_1,...,u_j, w_1,..., w_k$ are given for known $j,k,a,b$ which are related to the convergence  order, here chosen as fourth order. The total number of nodes is denoted by $n = j + m + k$. This method is of  high order accuracy when the Green's function is regular. 

As $k\rightarrow 1$, i.e. Green's function  (\ref{axhmg}) is close to singular and exhibits the same singular behavior as  (\ref{axlg}) for Laplace's equation. Simple calculation shows that the singular behavior of the normal gradient of $G^\chi$ is also the same as that of $G^0$. We add and subtract the singular Laplace kernel to obtain
\begin{align} \label{Gchi}
G^{\chi}(z,z_0,r,r_0)=\frac{k}{2\pi(rr_0)^{1/2}}\int_0^{1}\frac{{\rm{exp}}\left(-\Lambda\left[1-k^2t^2\right]^{1/2} \right)-1}{\left(1-k^2t^2\right)^{1/2}} \frac{dt}{(1-t^2)^{1/2}} + G^0(z,z_0,r,r_0),
\end{align}
so that the first term is regular, and we use the hybrid quadrature method as described above. Meanwhile, the singularity in $G^0$ is treated in a standard way, via  Gauss-log quadrature ~\cite{WangPapa2011JFM,Wang2012}. Expressions for the gradient of the Green's function  are given in Appendix~\ref{Ggrad} and the method of treating the singularity in derivatives of $G^\chi$  is the same as for $G^\chi$. Furthermore, in~Appendix~\ref{Ggrad} we  provide sample calculations of both the Green's function and its derivatives, which demonstrates the accuracy and  of our numerical method. We note that for thin Debye layers  $\Lambda \gg 1$, and the main contribution to the  integral in  (\ref{Gchi}) is localized near $t=1$ and $k=1$. Resolution studies show that our computation of the Green's function  has error of about $10^{-6}$   for  $\Lambda$ up to   $1414$, when $N = 2048$ is employed in the Alpert quadrature.

\subsection{Numerical procedure}
In this paper, we  focus on the deformation of axisymmetric drops. The azimuthal part of the surface integrations in each of the integral equations is carried out analytically, except for the ones with Green's functions from the modified Helmholtz equations, for which Alpert quadrature is implemented as described above. The drop interface is discretized by $N+1$ points, which divide it into $N$ segments.  The discretized equations assume the unknown `densities' $\phi_s$ and $\bm{u}$ vary linearly between node points along the interface. On each boundary element, this  gives  an integral  of the product of a linear (density) function and the Green's function.  Integrations of this product are carried out in double precision using six-point Gaussian quadrature  when the element is regular. As $\bm{x}_0\rightarrow \bm{x}$ the integrand is logarithmically singular, and Gauss-Log quadrature is used to handle the singularity (see also the implementation in~\cite{Wang2012}).   The normal and curvature along the drop interface are calculated by fitting cubic splines, which is similar to~\citet{StoneLeal90JFM}. 

\begin{figure}[!b]
\centering
\includegraphics[scale=0.65]{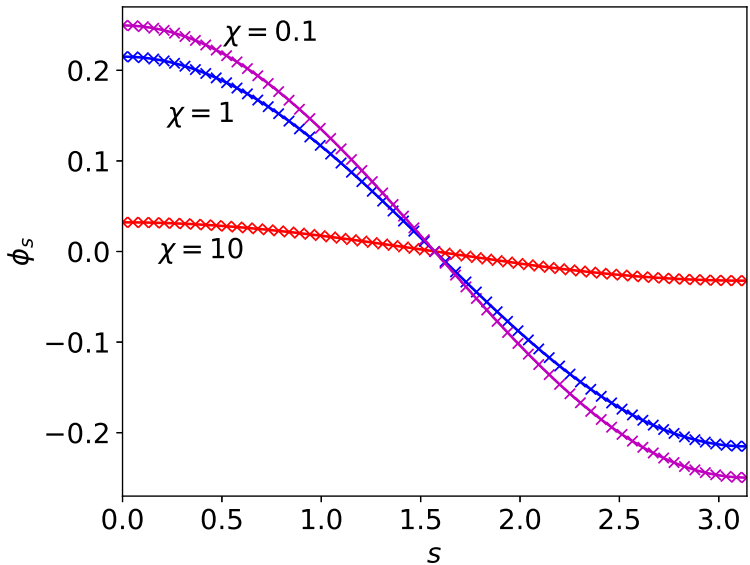}
\caption{Comparison  of numerically computed results (solid lines) for the surface potential $\phi_s$ and analytical solutions (cross symbols) for a spherical particle, from~\cite{HLK10}. Here $Q=10$ and $\chi$ is indicated in the figure. } \label{test1}
\end{figure}

The linear system that results from discretizing an integral equation is solved by using the Fortran subroutine $\mathrm{dgesv}$ in LAPACK. After obtaining surface velocities, the drop interface is advanced by Euler's method via the  kinematic condition.  The full method is second order accurate in space and first order accurate in time. 
A solution is deemed to be in a steady state when  max$|u_n|<10^{-4}$ along the drop interface.  For the steady state calculations reported here, $N=40\sim 70$ is typically enough to resolve the interface. The code for the Stokes droplet without an electric field has been extensively tested and used in earlier work~\cite{BPSW2013,WSB2014}. When  electrostatic fields are included in both the drop and surrounding phases in  the simpler case of $\chi=0$,  the code has been validated against small deformation theory as well as the results in~\cite{Sherwood88,LacHomsy07}, where good agreement is obtained. 
For example, in Figure~\ref{test1} we compare the numerically computed  electric potential $\phi_s$  for a spherical drop with an  analytical prediction from~\citet{HLK10}, for a uniform imposed field with  $\phi \rightarrow -z$ as $\bm{x} \rightarrow \infty$. Parameter values are  $Q=10$ and  $\chi=0.1,1,10$. Excellent agreement is obtained.

If the drop deforms into a highly elongated spheroidal shape or a spindle shape with conical ends, an adaptive regridding scheme is employed. In particular, grid points are redistributed using cubic spline interpolation to be inversely proportional to local curvature, so that the density of points is high near conical ends. One check on the overall method is to compare our calculated results with those in~\citet{Sherwood88} (e.g. their Figure 3). The results are in excellent agreement (see also our Figure~\ref{largeDf}).
If the drop becomes elongated  and exhibits end pinching or other breakup modes, larger $N$ is used (typically $N \sim 160$  to  $320$) together with adaptive time stepping. No special adaptive spatial regridding is applied in this case.

\section{Results and discussion}\label{sec:results}

We focus the discussion on the case where the interior and exterior fluids have equal viscosity, i.e., $\lambda=1$. The third term in equation (\ref{bie_u2}) is then absent, which greatly simplifies the numerics, but also allows for a rich bifurcation diagram and wide variety of unsteady shapes. Results for $\lambda\ne 1$ will be reported elsewhere.\\

\underline{\em Parameters in experiments}\\

{Microfluidic drops as small as 1$\mu$m are routinely generated and manipulated in experiments, although studies on drop deformation and  breakup commonly use millimeter-sized drops. Representative values of parameters in  experiments are: $\mathscr{E} \simeq 10^5$ V/m, $D\simeq 10^{-9}$ m$^2$/s, $\mu \simeq 100 Pa \cdot s$ ~\cite{SV2010, Saville1997} and $\gamma\simeq 10$ mN/m ~\cite{KDT2014}.
Assuming $R \simeq 1$mm  and approximating $\epsilon_2  \simeq 80 \epsilon_0$ (the permittivity of water), the electric capillary number at breakup is estimated as $E_b=\epsilon_2\mathscr{E}^2R/\gamma\simeq 10^{-1}$, consistent with the numerical results below. 
The conductivity $\sigma$ in our electrokinetic model is related to the ion density by $\sigma = 2e^2Dc^0/(k_BT)$,  assuming a  symmetric 1:1  electrolyte.
Using a representative bulk ion concentration $c^0 \simeq 10^{-7}$  moles/liter gives a conductivity $\sigma \simeq 10^{-9}$ S/m~\cite{Saville1997} with corresponding Debye layer thickness $\ell_D \simeq 10^{-6}$ m,  so that $\chi \simeq 10^3$.  A poorly conducting drop is obtained by specifying a thick Debye layer  $\chi  \lesssim 1$,
although care is required to insure that the equilibrium assumption  in (\ref{thermoeq}) is satisfied. This necessitates  that  the  charge relaxation time scale $t_e$  be much shorter than the  time scale for fluid motion $t_\mu$.  
Taking  $t_e=R^2/(\chi^2 D)$ and   $t_{\mu}=\mu/(\epsilon_2 \mathscr{E}^2)$   (which is a characteristic time for the Maxwell stress to deform the drop), we obtain   $t_e/t_\mu=Pe E_b/\chi^2$, where the Peclet number  $Pe=\gamma R/(\mu D)$.  
Using the above parameters, we find $t_e/t_\mu<<1$ for a millimeter drop when $\chi>>1$, but when $\chi \simeq 1$  it is necessary for  $R \lesssim 10 \mu $m to satisfy the condition on time scales. The latter estimate shows that the equilibrium assumption  is consistent with a Debye layer thickness on the order of drop size only  for a $10 \mu$m or smaller drop.
Finally, the dimensionless potential is estimated as $\beta \simeq 10^3$.  This suggests that $|\beta \phi_1|$  may not be small,  except in the perfect conductor limit $\chi>>1$  or $Q>>1$ in which case $\phi_1<<1$ (see e.g., Appendix  \ref{smd} ).  However, as noted, we will  apply the Debye-H\"{u}ckel approximation, even when it may not be formally valid.   }




\subsection{Steady state drops}
In this section we show the computed steady states of drops in a uniform imposed electric field. We note that unless specified otherwise, the reported simulation time is  rescaled  following \cite{Acrivos78jfm}  as $t=\tilde{t}\gamma/2\pi R\mu(1+\lambda)$, where $\tilde{t}$ is dimensional time. Following other work we measure drop deformation using the {Taylor deformation} parameter
\begin{align}
D_f = \frac{l-b}{l+b},
\end{align}
where $b$ and $l$ are {semi-axes of the drop at steady state perpendicular to and along the applied electric field, respectively.} 
To compute the steady  response curve we use continuation in the  parameter $E_b$: once a steady state solution  is obtained, we increase $E_b$ to a larger value and use the steady solution at the previous $E_b$ as initial data.

\begin{figure}[!ht]
\centering
\includegraphics[scale=0.45]{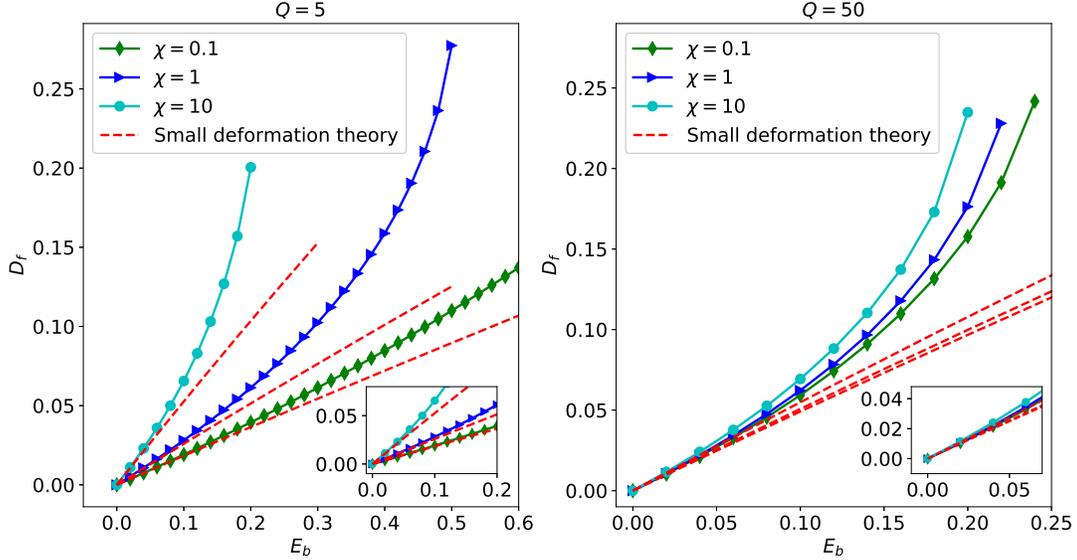}
\caption{Comparison between computed results (solid lines with filled symbols) and small deformation theory from \cite{HLK10} and Appendix~\ref{smd}
(dashed lines) with $Q$ and $\chi$ indicated in the figure. Insets: Magnification of results near $E_b=0$.}\label{dftest1}
\end{figure}

Figure~\ref{dftest1}  compares the drop deformation for fixed permittivity ratios $Q=5$ and $Q=50$ and a range of $\chi$. The deformation curves follow small deformation theory (see (\ref{smdf}) in Appendix~\ref{smd}) when $E_b$ is relatively small. The deformation is seen to be greater when $\chi$ is larger, with the same imposed  electric field $E_b$. 
This is because capillary pressure has a reduced effect, relative to electrostatic stresses, as $\chi$ is increased, per  (\ref{dFe}). This permits a more deformed surface before a local force balance between the capillary force and Maxwell (electrostatic) traction is reached.  Increasing $Q$ also tends to increase the deformation at a fixed imposed field strength.

\subsubsection*{\underline{Conducting drops}}

\begin{figure}[!b]
\includegraphics[scale=0.6]{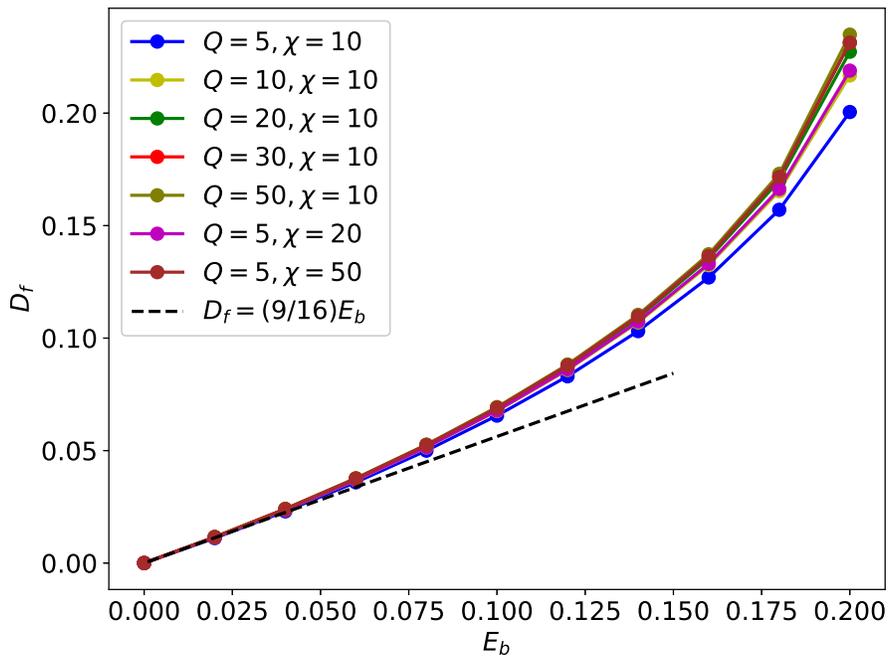}
\caption{Steady state deformation curves for various  $Q$ and $\chi$, corresponding to a highly conducting drop, {compared with the small deformation result $D_f=(9/16)E_b$}.}\label{dftest2}
\end{figure}

In our model, there are two ways to approach the conducting drop limit: either $\chi\rightarrow\infty$ or $Q\rightarrow \infty$. The surface potential $\phi_s$ in either case tends to zero, as can be seen
by taking the appropriate limit in the small deformation theory (see Appendix \ref{smd}). This  theory also shows that the deformation for  $\chi \gg 1$ and $Q  \gg 1$ is given by  $D_f \approx (9/16) E_b$, for $E_b \ll 1$. Figure~\ref{dftest2} shows the steady deformation curves for various values of $Q$ and $\chi$ which all correspond to  a highly conducting liquid drop. As expected, the deformation curves nearly overlap  each other. The critical value of $E_b$ at which point steady solutions no longer exist is roughly the same for each branch and is  about $0.21$, which is consistent with the value reported for a perfectly conducting drop in~\citet{KDT2014} and~\citet{DubashMestel07}.  The maximum interface potential over all the steady solutions represented in the figure is less than $0.03$.

\subsubsection*{\underline{Dielectric drops}}

\begin{figure}[!b]
\includegraphics[scale=0.8]{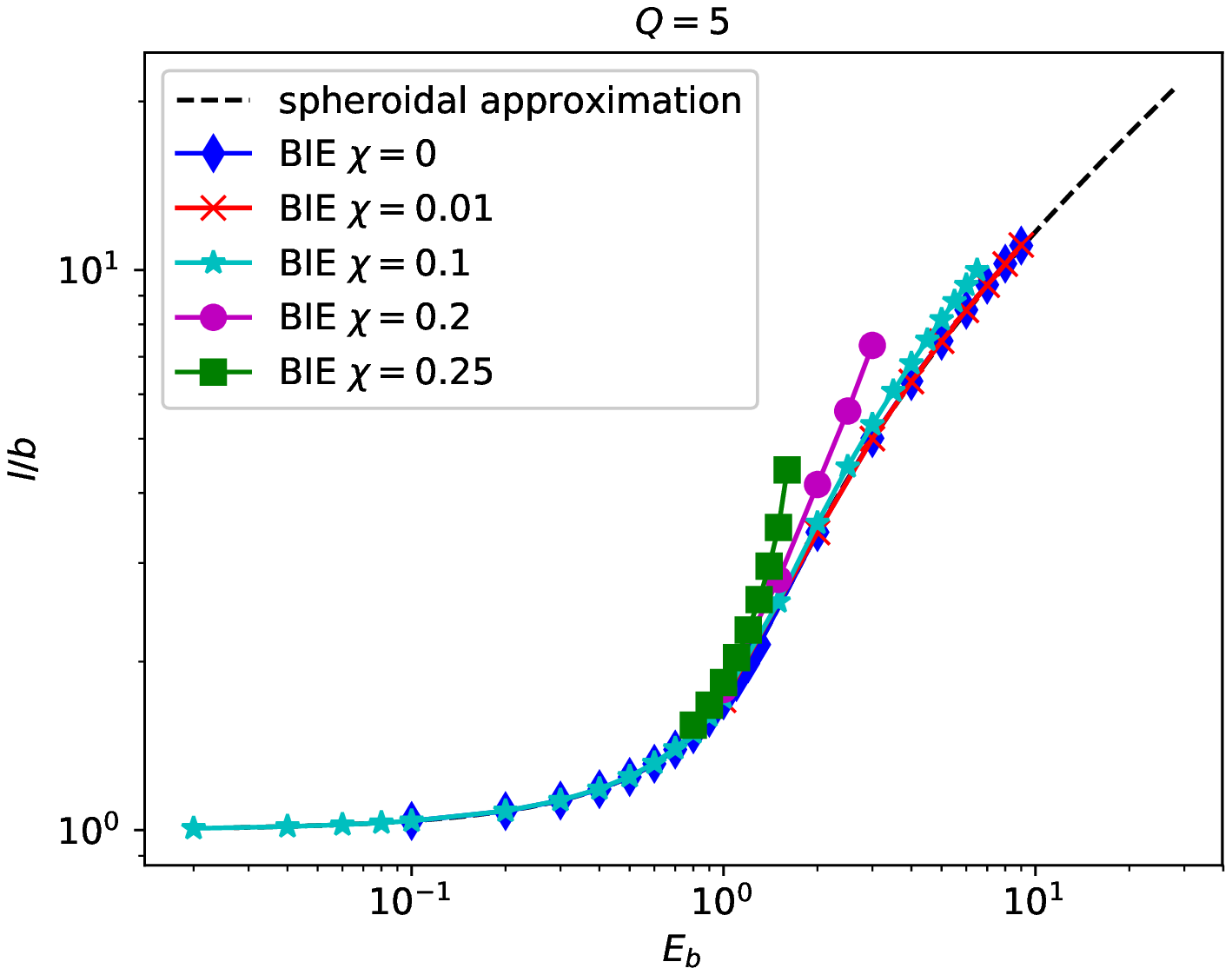}
\caption{Comparison of drop deformation for various $\chi$ and $Q=5$ together with the analytical results based on a spheroidal approximation for $\chi=0$.}
\label{largeDf}
\end{figure}

\begin{figure}[!h]
\includegraphics[scale=0.53]{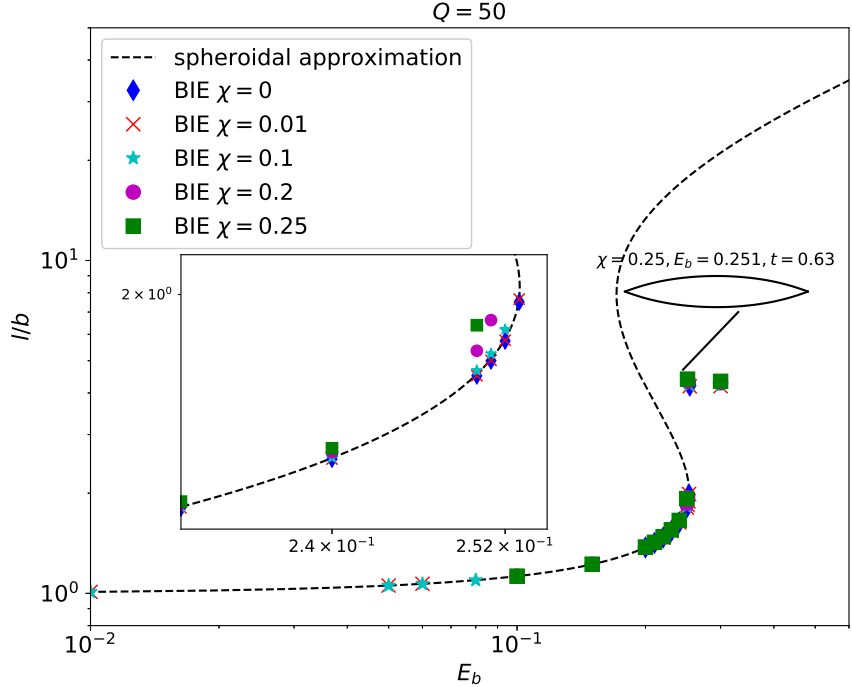}
\caption{Steady drop deformation for $Q=50$ and various $\chi$.   Theoretical steady-state response curve  using a spheroidal approximation for $\chi=0$ 
\cite{Saville1997, Pillai15} 
is  shown by a dashed-curve. The subplot provides  a zoomed-in view of the numerical data near the first turning point. The drop profile  for $E_b$ greater than the turning point on the lower branch exhibits unsteady pointed ends. 
}
\label{Q50}
\end{figure}

\begin{figure}[!t]
\includegraphics[scale=0.5]{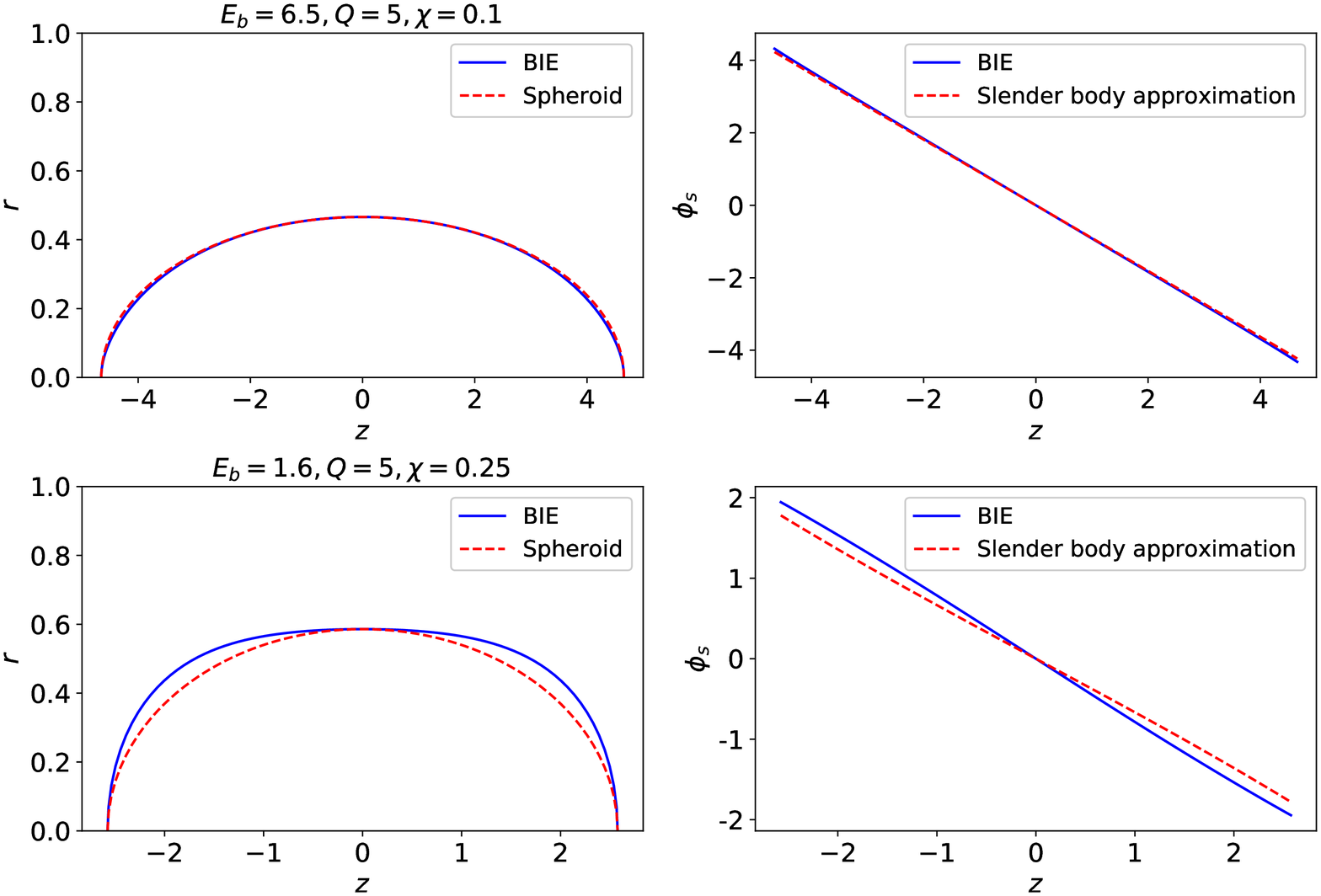}
\caption{Comparison of drop shape with a spheroid that has the same aspect ratio. Top left: aspect ratio $l/b\approx 10$ for a slightly conducting
drop with $\chi=0.1$, $E_b=6.5$, $Q=5$. Bottom left: aspect ratio $l/b\approx 4.4$ with $\chi=0.25$, $E_b=1.6$, $Q=5$.  Right: comparisons between calculated surface potentials and slender body approximation, for same parameter values as in  left panels.}
\label{pecompr2}
\end{figure}

When $\chi$ is small and $Q$ is not too large (i.e., $\chi \lesssim 1$ and $Q \lesssim 10^1$) , the drop is close to a perfect dielectric  suspended in an insulating medium (e.g., see the case $\chi=0.1$ and $Q=5$ in the left panel of Figure~\ref{dftest1}).   {We caution that time-dependent solutions to our model are  unphysical in this perfect dielectric limit, as noted in Section \ref{sec:results},  since the ratio of charge relaxation time scale to time scale for fluid motion, $t_e/t_\mu$, is not small.
Nevertheless,  comparison of the simulations  with the theoretical steady response curve provides a useful validation of the numerics.} Steady solution branches for such nearly dielectric drops are shown in  Figure~\ref{largeDf}, which   extends the plot in Figure~\ref{dftest1} to smaller values of $\chi$ and larger imposed field strength $E_b$. Instead of plotting the deformation as in Figure~\ref{dftest1}, we plot the aspect ratio {\red $l/b$} to which is better suited to the  wide range of $E_b$ used here.
We also overlay both the analytical solution using the  spheroidal approximation at $\chi=0$ and the boundary integral solution for a perfect dielectric drop in an insulating medium (the analytical expression for the spheroidal approximation is available in \citet{Pillai15}). It is seen that the computed deformation curves for $\chi=0$ and $0.01$ almost exactly lie on top of the analytical curve. For $\chi=0.1$, differences between the curves only occur when $E_b$ is sufficiently large. Consistent with previous observations at  small deformation, increasing $\chi$ promotes larger deformation for a given $E_b$. For $\chi=0.2$ and $0.25$, deviation from the  insulating drop limit occurs at smaller $E_b$, as expected. An aspect ratio of $l/b\approx 4.4$ is quickly reached at about $E_b=1.6$ for $\chi=0.25$,  beyond which the drop is found to be unstable and steady solutions no longer exist.

{For larger $Q$ (e.g., $Q = 50$ in Figure \ref{Q50}) 
the spheroidal approximation gives an 'S'-shaped curve  ~\cite{Saville1997, Pillai15}. The steady solutions computed  by our time-dependent simulations converge to the lower branch of the theoretical response curve   as $\chi\rightarrow 0$. 
Using our model, it was not possible to capture the jump to the upper branch of the deformation curve. Instead, for $E_b$ greater than the critical value $E_b \simeq 0.25$ at
the  turning point on the lower branch, an unsteady pointed drop develops and the numerics eventually break down, as discussed more fully in Section \ref{sec:breakup}. This is  similar to the results of 
\cite{Sherwood88} using a dielectric model, but contrasts with the time-depedent simulations of 
~\cite{Pillai15} using a full electrokinetic model, which evolve to steady solutions on the upper branch. The latter discrepancy may be due to the equilibration time for the Debye layer charge, which is here assumed to be fast, but can evolve more slowly in the model of ~\cite{Pillai15}.}

It is natural to carry out a slender body analysis when a highly elongated drop is obtained. In Appendix~\ref{slendapp}, we assume a highly deformed drop and obtain an asymptotic approximation of the electric field for a spheroidal drop, see (\ref{e1d}). The field is shown to be almost uniform.  Our results serve as a correction to the result in~\cite{SLB99} by taking into account the presence of ions. Integration of the  field  gives the electric potential, and the theoretical drop shape and electrostatic potential  are compared to our numerically computed  solutions in  Figure~\ref{pecompr2}. In the upper panels of the figure, the drop is highly elongated with aspect ratio about $10$ (i.e. slenderness parameter $\epsilon=b/l\approx 0.1$). Both the shape and electric potential are in excellent agreement with theory. In the bottom panels, the drop shape  is shown to deviate from a spheroid with the same aspect ratio (about $4.4$). The interfacial potential is also slightly different from the prediction of slender-body analysis but the agreement is still reasonably good.

\subsection{Breakup behavior}   \label{sec:breakup}
In this section, we investigate drop deformation for parameter values in which steady state solutions do not exist.  Several different types of unsteady solution  are observed (depending on parameter values), which are classified  into three groups: (i) conical end formation, (ii) end-splashing, and (iii) open end stretching:  A few case studies are presented before a summary is given. 

\subsubsection*{\underline{Conical end formation}}
\begin{figure}[!t]
\includegraphics[scale=0.5]{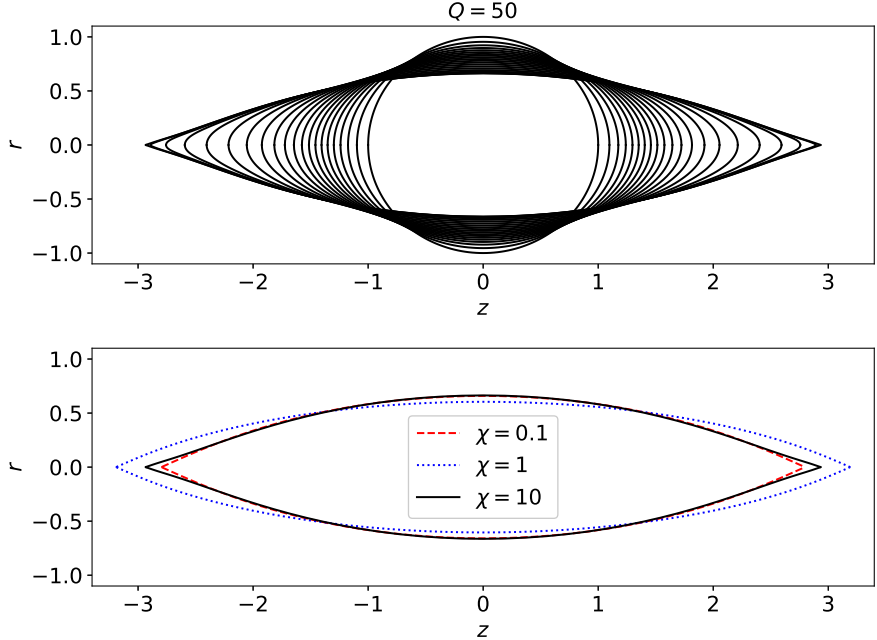}
\caption{Upper panel: Evolution of drop with $Q=50$, $\chi=10$ and $E_b=0.26$. Bottom panel: Drop shapes at breakdown of the numerical scheme for $Q=50$ and $E_b=0.26$.}\label{cusp}
\end{figure}

In Figure~\ref{cusp}, conical end formation is shown for conducting drops with $Q=50$ and $\chi=0.1,1,10$. In the upper panel of  the figure, a time-sequence of unsteady drop shapes are shown  for $\chi=10$, $Q=50$, starting from an initially spherical shape. In the lower panel, drop shapes at the point at which the simulation is terminated are shown for $\chi=0.1,1,10$, which all show the  formation of  unsteady  conical drop tips that are similar to the shapes reported in~\cite{Sherwood88,SLB99,BFKV2006}.   The simulation is terminated when the tip curvature $\kappa_{tip}$ becomes sufficiently  large that the number of  grid points and time-step required to resolve the interface make the simulation too computationally costly.   It is argued in Appendix~\ref{conical}, based on equation  (\ref{cuspBD}),  that conical end formation can occur only for sufficiently large $Q$, or more precisely $Q \gtrsim15$, regardless of  $\chi$.

\begin{figure}[h!]
\includegraphics[scale=0.75]{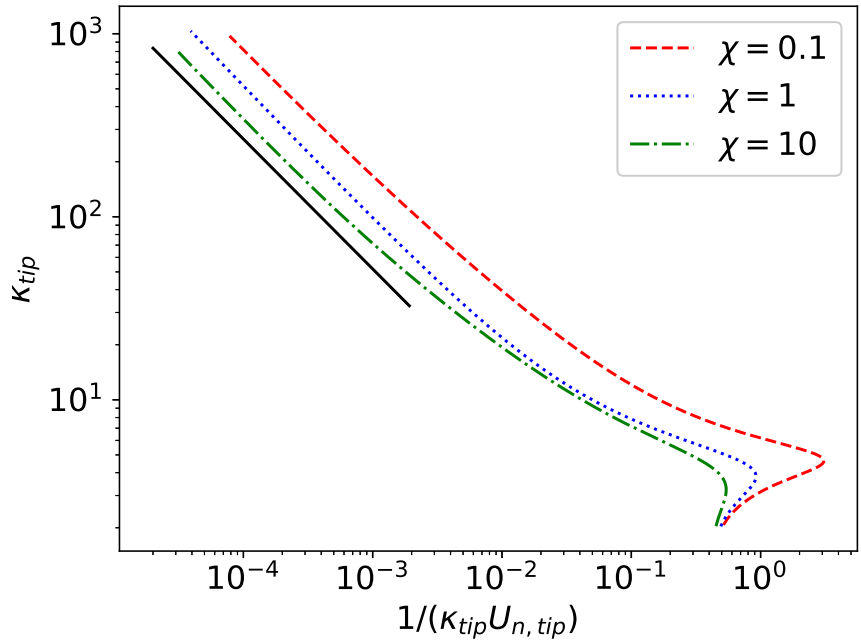}
\caption{{Evolution of interface tip curvature $\kappa_{tip}$ versus $1/(\kappa_{tip}U_{n,tip})$ for the three cases in Figure~\ref{cusp}.  Here, $U_{n,tip}$ is the normal velocity at the tip. Following the scaling law $\kappa_{tip} =O(\tau^{-\delta})$ and $U_{n,tip}=O(\tau^{\delta-1})$ from \citet{FKV2008}, we plot $\kappa_{tip}$ versus the time to  singularity $\tau \sim 1/(\kappa_{tip} \cdot U_{n,tip})$. The estimated or average slope $\delta \approx 0.71$ is shown as a black solid line.
}}\label{fcurv}
\end{figure}

\citet{FKV2008} present an analysis of conical singularity formation for a charged, perfectly conducting drop  in an insulating medium. They find that the singularity formation is self similar with $\kappa_{tip} =O(\tau^{-\delta})$ and $U_{n,tip}=O(\tau^{\delta-1})$, where $\tau=t_s-t$ is the time to singularity formation (i.e., the singularity occurs at $t=t_s$) and  $\delta$ is a similarity exponent that depends on the opening angle of the cone. Although our model incorporating electrokinetic effects in the drop interior via the linearized Poisson-Boltzmann equation is different from \citet{FKV2008}, this form of the similarity scalings seems to remain unchanged based on our numerical results. We will determine $\delta$ from numerical data, but one difficulty in doing so is that the singularity time $t_s$ is unknown. While it can be estimated from numerical data, we take a different approach.  Assuming the above self-similar scalings,
then $\tau \sim 1/(\kappa_{tip} \cdot U_{n,tip})$, and since  accurate values for  $\kappa_{tip}$ and $U_{n,tip}$ are provided by the numerical data, we can replace $\tau$ by $1/(\kappa_{tip} \cdot U_{n,tip})$ in a log-log plot to determine $\delta$.  Such a plot of the time evolution of tip curvature $\kappa_{tip}$  versus $1/(\kappa_{tip} \cdot U_{n,tip})$ for a drop which forms conical tips  is shown in Figure~\ref{fcurv}.  The figure shows  linear behavior for 
$\log \kappa_{tip}$ with a slope that is {very slightly dependent on} $\chi$. We estimate the slope magnitude or similarity exponent to be very near {$\delta = 0.71$}, which is close to the value of $0.72$ reported in  \citet{FKV2008}. Our simulations give slightly different results for the cone 
angles than~\citet{BFKV2006} and \citet{FKV2008}. We find the semi-angles are between $21\sim 24$ degrees for the different $\chi$ values. In the work of~\citet{BFKV2006}, the semi-angle is shown to be dependent on the viscosity ratio and  is about $25$ degrees for $\lambda=1$. A slightly different result of about $27.5$ degrees is reported in~\citet{FKV2008} (for a different model). While we cannot completely rule out numerical error as a source of  the variation of cone angle with $\chi$ found here, resolution studies suggest that the computed angles are well resolved. We note that the current numerical method is only able to resolve about  $2$ decades of scaling in the space-time neighborhood of the singularity, which is similar to the other cited studies. For much more than this, and for a more detailed investigation of cone angles,  it is anticipated that a specialized numerical treatment of the emerging singularity is needed.

\subsubsection*{\underline{End-splashing mode}}

\begin{figure}[!ht]
\includegraphics[scale=0.53]{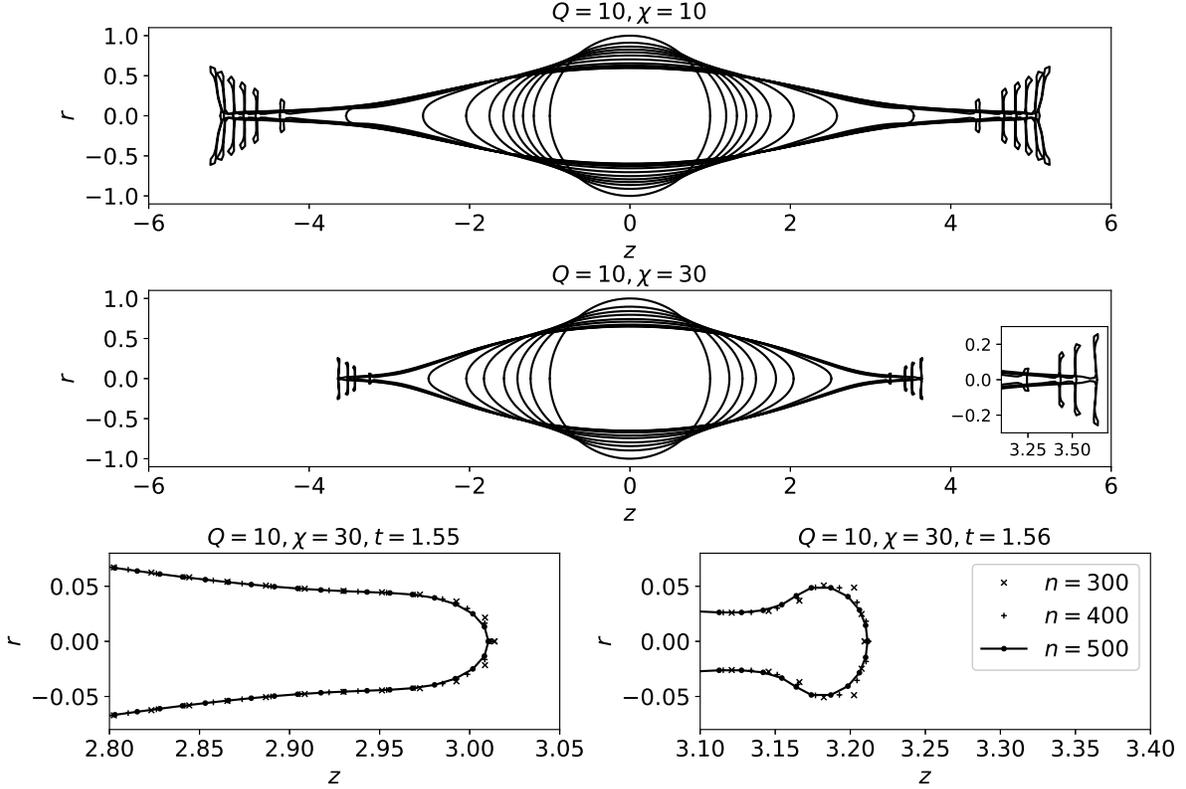}
\caption{Breakup of a viscous drop for $Q=10$ and $E_b=0.26$ with $\chi=10$ in upper panel ($t=0, 0.25, 0.49, 0.74, 0.98, 1.23, 1.47, 1.66, 1.77, 1.83, 1.86, 1.88, 1.90, 1.92, 1.94$) and $\chi=30$ in the middle panel ($t=0, 0.30, 0.60, 0.90, 1.20, 1.35, 1.50, 1.56, 1.58, 1.59, 1.61$). 
The inset of the middle panel show local finger formation before breakup for $\chi=30$. The lower two panels show  tip profiles at different resolution $N$ for $\chi=30$ at times before (left) and after the snail head is formed (right).  The  profiles are well-resolved,  at least up to the onset of snail formation.  
}\label{tipb}
\end{figure}

When $\chi$ is large enough (see Figure~\ref{phase0_5} for precise values),   conical end formation is replaced by  a small finger that is  emitted from the tip, nearly perpendicular to the axis of symmetry or  $z$-axis. This behavior persists even in  the highly conducting drop limit of large $\chi$.  The interface shape near the ends eventually  evolves into a 'snail-head' that forms in the vertical direction. We call this the `end-splashing mode'. Representative examples of end-splashing are  plotted in the top two  panels of Figure~\ref{tipb}. For these simulations, $N=320$ and the profile is  well resolved  up to the point when bulbous ends first form, which marks the onset of snail-head finger formation (see bottom panel). Resolution studies of the fully developed snail head profile, e.g. at the final time in the upper panel of Figure~\ref{tipb}, show similar shapes, but slightly decreased snail head length, as resolution is increased. In a 3D view, the drop end looks like a disk or nearly flat cone with a ring rim. This is similar to the so-called dimple formation and lobe-breakup solution reported in~\citet{KDT2014} (see their figure 11 for example). This interface morphology is distinctly  different from that observed in~\citet{Taylor64,GB2005,BFKV2006} and \citet{FKV2008}, where a Taylor-cone-like solution  first develops, then  is followed by the ejection of a thin fluid thread in the axial direction.

{The end-splashing breakup mode has not been observed in simulations of  the leaky dielectric model, although 
for highly conducting fluids}, similar  breakup behavior has been reported in experiments and  simulations of a perfect conductor model in \citet{KDT2014}.  The experiments have $NaCl$ added to the drop phase, suggesting that ions in the drop may contribute to the fingering instability. More recently, ~\citet{MLHMG2016} observed a similar end-splashing mode in experiments and simulations using an electrokinetic model for conducting fluids in a pendant drop problem,
although there it was attributed to the effect of a more viscous fluid in the  surrounding medium. In the current study, this behavior is also obtained when the viscosity ratio is one.   The middle panel of Figure~\ref{tipb} shows a similar fingering instability  for the  larger value $\chi=30$, except that a narrower finger is formed. Our numerical results show a trend of decreasing  finger width with increasing $\chi$.

\subsubsection*{\underline{Open end stretching}}
\begin{figure}[!t]
\includegraphics[scale=0.65]{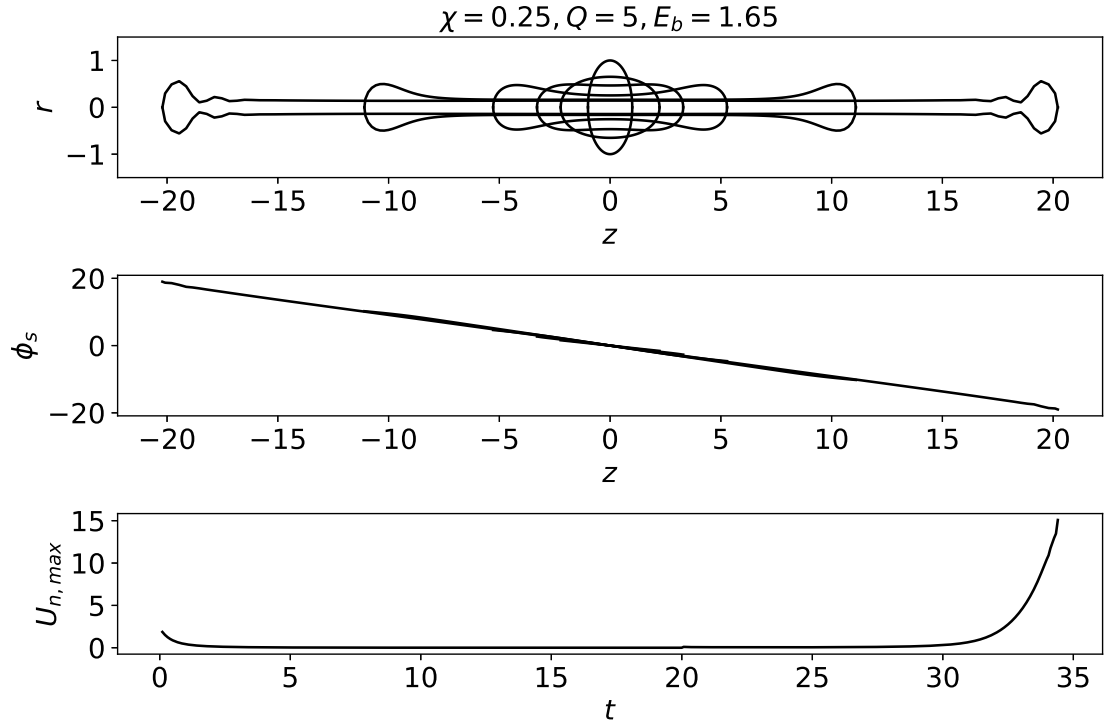}
\caption{Breakup of a viscous drop for $Q=5$ and $E_b=1.65$ with $\chi=0.25$. Top: Drop profiles at times $t= 0, 1.73, 29.31, 32.14, 33.56, 34.41$.   Middle: interfacial potential $\phi_s$ versus $s$, at the same $t$ as top. Bottom: maximum normal velocity versus $t$. }\label{long}
\end{figure}

When both $\chi$ and $Q$ are both moderate in size (roughly of the order $10^{0}$ to $10^{1}$)  our model exhibits relatively long drops, i.e., with $\epsilon=b/l<0.01$. One such example is presented in Figure~\ref{long} for $E_b$ slightly larger than the critical value for nonexistence of steady sates. The top two panels show the drop shape and interface potential, respectively,  while the  bottom panel plots the maximum normal velocity. As it evolves,  the drop forms a cylindrical central  thread  and the electrostatic field is  nearly  uniform and directed along the axis of symmetry.  During the evolution, the normal velocity decreases and the  drop nearly settles into an elongated steady state. However, the elongational velocity is reestablished after about $t\approx 30$ as the electric traction  overwhelms surface tension. 
Two `blobs' develop at the drop ends in a manner  similar to the initial stages  of the end-splashing mode in  Figure~\ref{tipb}. However, due to its highly elongated cylindrical thread, we classify this to be a third breakup mode, `open end stretching'.  The  simulation in Figure~\ref{long} was stopped when the drop aspect ratio exceeded $100$. 
{The open-end stretching found here is similar to `end-pinching' solutions computed using the leaky dielectric model (see, e.g., \cite{LacHomsy07, Sengupta2017}).  
We do not find multi-lobe end-pinching solutions, i.e.,  with internal circulation, like that in Figures 7 and 9 of \cite{Sengupta2017}   (see also \cite{LacHomsy07}). We believe these differences  may be  due to the absence of surface charge and tangential interfacial electric stress in our model. }

\subsubsection*{\underline{Summary of breakup modes}}
\begin{figure}[!b]
\includegraphics[scale=0.65]{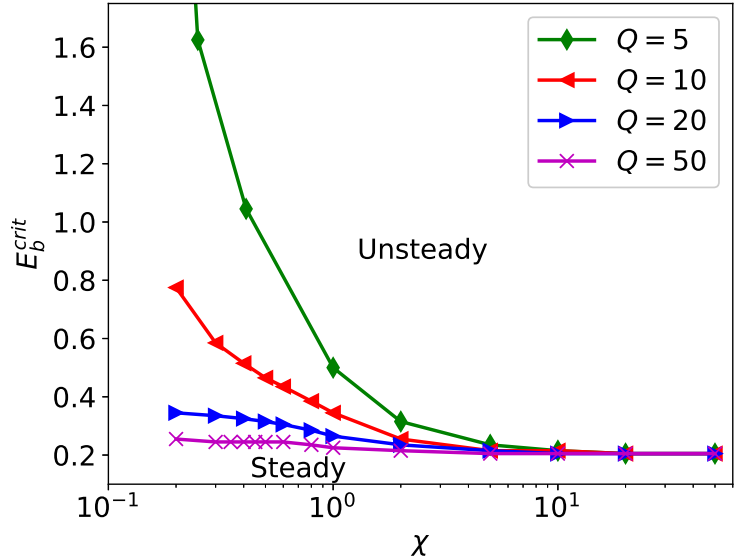}
\caption{Boundaries separating steady  ($S$-region) and unsteady ($U$-region) solutions in the  $E_b-\chi$ plane for various $Q$.}  \label{phase1}
\end{figure}

\begin{figure}[h!]
\includegraphics[scale=0.45]{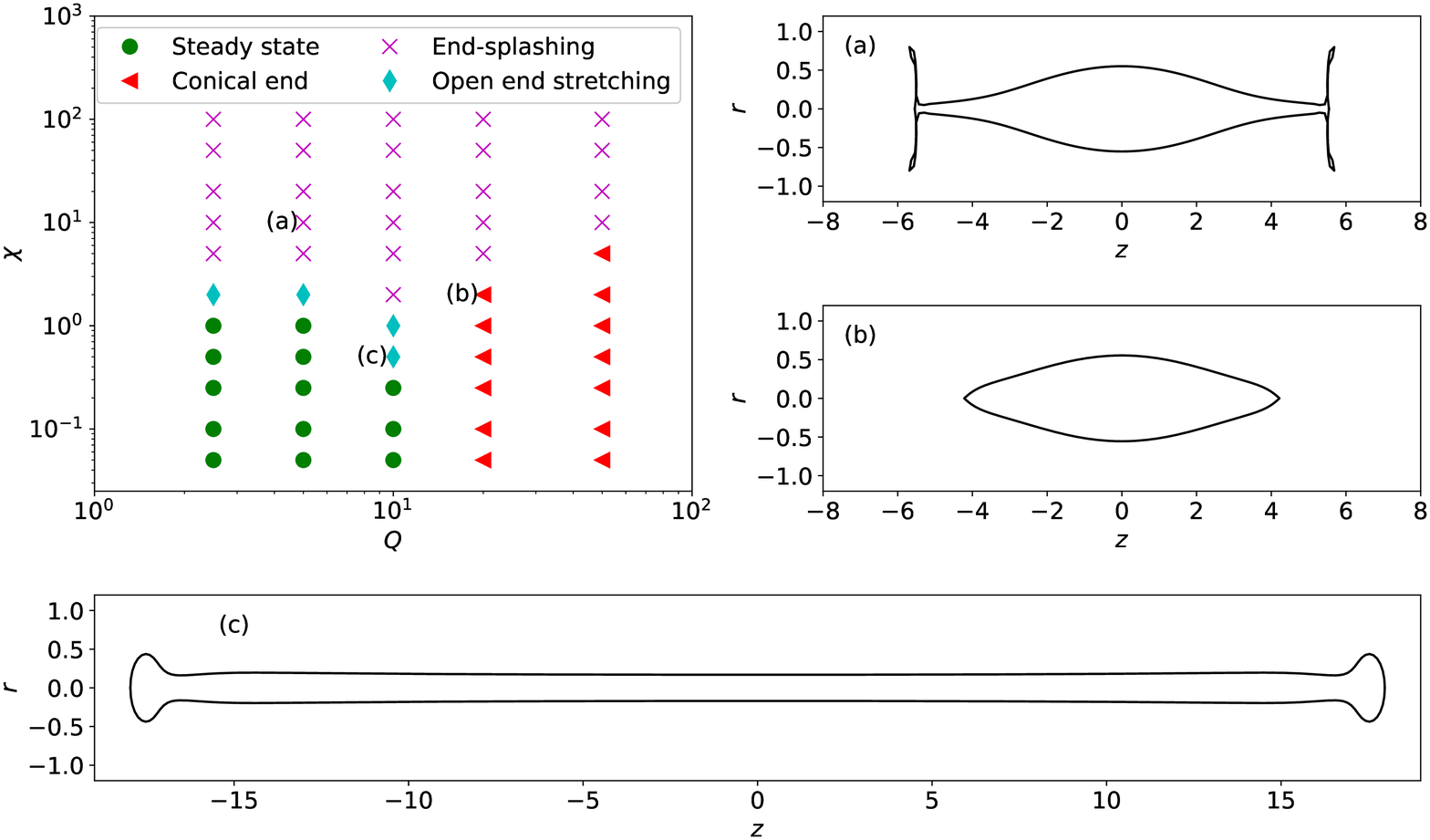}
\caption{ Phase diagram of steady shapes and breakup modes in $Q-\chi$ space with $E_b=0.5$.  Interface shapes at markers (a)-(c) on the phase diagram are shown in panels at right and below.  (a) End-splashing ($Q=5, \chi=10$), (b) conical end formation ($Q=20, \chi=2$) and  (c)  open end stretching ($Q=10, \chi=0.5$) }\label{phase0_5}
\end{figure}

We summarize our results in Figures~\ref{phase1} and~\ref{phase0_5}.   Figure~\ref{phase1} shows numerically determined curves in $E_b - \chi$ space  that separate  regions where steady drop shapes exist (or $S$-regions) from those with purely unsteady dynamics ($U$-regions).  Toward the conducting drop limit, i.e., for sufficiently large $Q$, the $U$-region is roughly independent of $\chi$ and  occurs above $E_b\approx 0.22$.  For moderate or small $Q$, a narrow $S$-region occurs when $\chi$ is sufficiently small, i.e.,  as the  perfect dielectric limit is approached. Alternatively, Figure~\ref{phase0_5} plots the phase diagram in $Q-\chi$ space with a fixed electric field strength $E_b=0.5$. The behavior in the $U$-region  is further classified by breakup mode.  When $Q \lesssim 15$,  we find the $S$-region  for small $\chi$ and  end-splashing breakup modes (Figure~\ref{phase0_5}(a)) for large $\chi$, while for $\chi$  in between we find open end stretching modes as shown at example point Figure~\ref{phase0_5}(c). 
When $Q \gtrsim 15$, conical end  solutions are found for  sufficiently small $\chi$ (Figure~\ref{phase0_5}(b)), consistent with the asymptotic theory in Appendix  \ref{conical}, whereas for larger $\chi$ we find end-splashing breakup modes.   



We conclude this section by  noting that only the three aforementioned breakup modes are observed in the current study when the viscosity ratio is unity.  {Other than end-splashing, solutions which develop topological singularities, i.e., where the drop fractures into two or more droplets,  are not seen in the electrokinetic  model considered here.  Preliminary computations show  that other  breakup modes can occur in our model for  $\lambda<1$, and will be reported elsewhere.} 

\section{Conclusion}\label{sec:concl}
We have developed  a robust and accurate numerical method to evaluate Green's functions for the linearized Poisson-Boltzmann equation  and  applied it to solve the moving boundary problem for Stokes flow,  including electrostatic forces. The method is used to investigate the steady deformation of an electrolyte drop  suspended in an insulating fluid medium, for viscosity ratio $\lambda=1$. We demonstrate that the conducting drop limit can be approached through either $Q\rightarrow\infty$ or $\chi\rightarrow\infty$, and in these limits, the electric field inside the drop vanishes. For large but finite $Q$,  however,  the electric field inside the drop is nonzero and can contribute to the deformation. The perfect dielectric drop limit  is  $\chi  \rightarrow 0$, and for small $\chi$ we find that  a highly elongated steady drop is possible. For given imposed field strength $E_b$, the presence of electrolyte enhances the deformation, and  increasing $\chi$ leading to more deformed drops. When the drop is long and slender, results from our  full numerical simulations agree well with approximate solutions based on slender body theory. Finally, we study  drop breakup behavior by choosing parameters in the regime where steady solutions no longer exist. In addition to conical end formation, we find other two breakup modes, which we call `end-splashing' and `open end stretching'. The type of break up  depends on parameter values 
 and an example phase diagram is  presented which illustrates the dependence  on two of those parameters ($Q$ and $\chi$), for fixed $E_b$.

{Finally, we note that the electrokinetic model and numerical method developed here can provide a framework for extensions that incorporate additional physics, including surface charge or electrolytic effects in the exterior fluid.} 

\acknowledgements{
MS acknowledges the financial support from National Science Foundation through the grant DMS-1412789, MM acknowledges the financial support from Natural Science Foundation of China (No. 11701428) and "Chen Guang" project supported by Shanghai Municipal Education Commission and Shanghai Education Development Foundation.
}

\appendix
\section{Computation of axisymmetric Green's functions and their derivatives }\label{Ggrad}
In this section, we present the derivatives of the  axisymmetric Green's functions (\ref{axhmg}) and (\ref{axlg}).   Gradients of the axisymmetric Green's function for Laplace's equation are given by  
\begin{align}
	&\frac{\partial G^0}{\partial z}=-\frac{(z-z_0)k^3E(k)}{8\pi (rr_0)^{3/2}(1-k^2)}, \\
	&\frac{\partial G^0}{\partial r}=-\frac{k^3}{8\pi(rr_0)^{3/2}}\left[\left(\frac{r-r_0}{1-k^2}-\frac{2r_0}{k^2}\right)E(k) + \frac{2r_0}{k^2}K(k) \right],
\end{align}
where $K(k)$ is the complete elliptic integral of the first kind (see (\ref{kfun})) and 
\begin{equation}
	E(k) = \int_{0}^{\pi/2}\sqrt{1-k^2\cos^2\theta}d\theta
\end{equation}
is the complete elliptic integral of the second kind. For the axisymmetric Green's function $G^\chi$ of the modified Helmholtz equation, we find that the Green's function gradient is given by 
\begin{align}
	\frac{\partial G^\chi}{\partial z}= \frac{k_z}{2\pi(rr_0)^{1/2}}\int_0^{\pi/2}&\frac{1+\Lambda\left(1-k^2\cos ^2 \theta \right)^{1/2}}{\left(1-k^2\cos ^2\theta \right)^{3/2}}\exp\left(-\Lambda\left[1-k^2\cos ^2 \theta \right]^{1/2} \right) d\theta, \\
	\frac{\partial G^\chi}{\partial r}=\frac{1}{2\pi(rr_0)^{1/2}}\int_0^{\pi/2}&\frac{1+\Lambda\left(1-k^2\cos ^2 \theta \right)^{1/2}}{\left(1-k^2\cos ^2\theta \right)^{1/2}}\left(\frac{k_r}{1-k^2\cos^2\theta}-\frac{k}{2r} \right) \nonumber\\
	&\times\exp\left(-\Lambda\left[1-k^2\cos ^2 \theta \right]^{1/2} \right) d\theta.
\end{align}
For the general calculations of Green's functions for Laplace equation, details can be found in~\citet{Poz2002}.

\begin{table} [h!]
\caption{$\chi = 0.1$, Gauss-Trapezoidal parameters $j=7$, $k=6$, $n = j+k+m$}\label{tab:A1}
\begin{tabular}{lc c c c c c c c c c c c c }
\hline \hline
\multirow{2}*{$n$}& & \multicolumn{5}{c}{Gauss-Trapezoidal} &     & \multicolumn{5}{c}{Gauss-Chebyshev}\\
\cline{3-7}\cline{9-13}
 & & $G^\chi$ & & $G_z^\chi$& &$G_r^\chi$ & &$G^\chi$ & & $G_z^\chi$& &$G_r^\chi$ \\
\hline
  16 && 1.95602152 && 207.50604021 && -208.60234657 && 1.95602514 && 207.50604303 && -208.60234683 \\ 
  32 && 1.95602152 && 207.50604021 && -208.60234657 && 1.95602243 && 207.50604261 && -208.60234833 \\ 
  64 && 1.95602152 && 207.50604021 && -208.60234657 && 1.95602175 && 207.50604218 && -208.60234839 \\ 
 128 && 1.95602152 && 207.50604021 && -208.60234657 && 1.95602158 && 207.50604176 && -208.60234809 \\ 
 256 && 1.95602152 && 207.50604021 && -208.60234657 && 1.95602154 && 207.50604134 && -208.60234770 \\ 
 512 && 1.95602152 && 207.50604021 && -208.60234657 && 1.95602153 && 207.50604093 && -208.60234730 \\ 
1024 && 1.95602152 && 207.50604021 && -208.60234657 && 1.95602153 && 207.50604057 && -208.60234693 \\   
     
\hline \hline
\end{tabular}
\end{table}

\begin{table} [h!]
\caption{$\chi = 1$}\label{tab:A2}
\begin{tabular}{lc c c c c c c c c c c c c }
\hline \hline
\multirow{2}*{$n$}& & \multicolumn{5}{c}{Gauss-Trapezoidal} &     & \multicolumn{5}{c}{Gauss-Chebyshev}\\
\cline{3-7}\cline{9-13}
 & & $G^\chi$ & & $G_z^\chi$& &$G_r^\chi$ & &$G^\chi$ & & $G_z^\chi$& &$G_r^\chi$ \\
\hline
  16 && 1.66552053 && 207.50556904 && -208.52289051 && 1.66588210 && 207.50585107 && -208.52291721 \\ 
  32 && 1.66552054 && 207.50556904 && -208.52289050 && 1.66561085 && 207.50580881 && -208.52306613 \\ 
  64 && 1.66552056 && 207.50556904 && -208.52289048 && 1.66554306 && 207.50576654 && -208.52307180 \\ 
 128 && 1.66552057 && 207.50556904 && -208.52289048 && 1.66552612 && 207.50572433 && -208.52304161 \\ 
 256 && 1.66552057 && 207.50556904 && -208.52289048 && 1.66552191 && 207.50568240 && -208.52300272 \\ 
 512 && 1.66552057 && 207.50556904 && -208.52289048 && 1.66552087 && 207.50564156 && -208.52296267 \\ 
1024 && 1.66552057 && 207.50556904 && -208.52289048 && 1.66552063 && 207.50560466 && -208.52292600 \\ 
    
\hline \hline
\end{tabular}
\end{table}

\begin{table} [h!]
\caption{$\chi = 10$}\label{tab:A3}
\begin{tabular}{lc c c c c c c c c c c c c }
\hline \hline
\multirow{2}*{$n$}& & \multicolumn{5}{c}{Gauss-Trapezoidal} &     & \multicolumn{5}{c}{Gauss-Chebyshev}\\
\cline{3-7}\cline{9-13}
 & & $G^\chi$ & & $G_z^\chi$& &$G_r^\chi$ & &$G^\chi$ & & $G_z^\chi$& &$G_r^\chi$ \\
\hline

  16 && 1.12249190 && 207.47248568 && -208.10737309 && 1.15507371 && 207.50023604 && -208.11656235 \\ 
  32 && 1.12249346 && 207.47248566 && -208.10737197 && 1.13128003 && 207.49634418 && -208.12532268 \\ 
  64 && 1.12249500 && 207.47248564 && -208.10737086 && 1.12472887 && 207.49220608 && -208.12550479 \\ 
 128 && 1.12249602 && 207.47248562 && -208.10737012 && 1.12305031 && 207.48800774 && -208.12247820 \\ 
 256 && 1.12249628 && 207.47248562 && -208.10736994 && 1.12262964 && 207.48382016 && -208.11859296 \\ 
 512 && 1.12249621 && 207.47248562 && -208.10736999 && 1.12252599 && 207.47973719 && -208.11458941 \\ 
1024 && 1.12249620 && 207.47248562 && -208.10737000 && 1.12250164 && 207.47604760 && -208.11092268  \\
 \hline \hline
\end{tabular}
\end{table}

In table~\ref{tab:A1} to \ref{tab:A3}, we show data from a computation  of $G^\chi$ comparing the  Gauss-Chebyshev method  and Gauss-Trapezoidal method (or Alpert quadrature), for $\chi=0.1,1,10$,  and different $n$ or number of  quadrature points.  The points of evaluation are  $(z_0, r_0)=(\cos(\pi/4), \sin(\pi/4))$,  $(z, r) = (\cos(\pi/4+\pi/4096), \sin(\pi/4+\pi/4096))$, so that $(z,r)$ is slightly different  from $(z_0,r_0)$.
When $(1-t^2)^{-1/2}$ is treated as a weight function in (\ref{axhmg1}), the integration, for a regular function $f(x)$, can be done by Gauss-Chebyshev quadrature
\begin{align} \label{eq:GCQ}
\int_{-1}^1 \frac{f(t)}{(1-t^2)^{1/2}}dt=\frac{\pi}{n}\Sigma_{j=1}^n f(t_{j,n})+\frac{2\pi}{2^{2n}(2n)!}f^{(2n)}(\eta), 
\end{align}
for some $-1<\eta<1$ and
\begin{equation}
t_{j,n}=\cos\left(\frac{2j-1}{2n}\pi\right).
\end{equation}

For $\chi=0.1$, both quadrature methods work well, however, Gauss-Trapezoidal quadrature is more accurate than Gauss-Chebyshev for moderate and large $\chi$, as seen in Table~\ref{tab:A2} and \ref{tab:A3}. In particular, when $\chi=10$, it is seen that convergence is poor for the Gauss-Chebyshev method. For example, $G^{\chi}_r$ obtains only one digit of precision at the largest $n$.
In this paper, we therefore employ Gauss-trapezoidal quadrature since it has performed well in our  tests. We have not performed an extensive investigation of quadrature methods, as this is beyond the scope of current paper.

\section{Small deformation theory}\label{smd}
Small deformation analysis of an electrolyte droplet immersed in a dielectric fluid  and deformed by a nonuniform electric field is provided by \cite{HLK10}, from which the following solution for a uniform field can be easily  recovered. 
When the drop is spherical, the general solution for the electric potential  is given in spherical radial and polar coordinates $\bar{r}, \bar{\theta}$ by
\begin{equation}
\phi_1= - A_1 ( \bar{r}) \cos(\bar{\theta}), \quad \phi_2 = -\left(1-\frac{A_2}{\bar{r}^3}\right) \bar{r} \cos(\bar{\theta})
\end{equation}\label{slnphi}
with
\begin{align}
& i_1(x)=\frac{x\cosh(x)-\sinh(x)}{x^2},\\
& i_2(x) = \frac{(x^2+3)\sinh(x)-3x\cosh(x)}{x^3},\\
& A_1(\bar{r})= \frac{3 i_1(\chi \bar{r})}{(Q+2)i_1(\chi)+Q\chi i_2(\chi)},\\
& A_2 = \frac{(Q-1)i_1(\chi)+Q\chi i_2(\chi)}{(Q+2)i_1(\chi)+Q\chi i_2(\chi)}.
\end{align}

When deformability is included, first order perturbation can be used to approximate the shape of the drop. Assuming the shape is perturbed slightly when $E_b\ll 1$, \cite{HLK10} derived an expression for the  deformation 
\begin{equation}
D_f \approx \frac{3E_bh(\chi,Q)}{4+E_b h(\chi,Q)}\approx \frac{3}{4}E_bh(\chi,Q) + O(E_b^2)\label{smdf}
\end{equation}
where
\begin{align}
h(\chi,Q)=\frac{1}{12Q}\left[(Q-1)(1+2A_2)^2+(\chi^2+1-Q)A^2_1(r=1)Q\right].
\end{align}
It is instructive to note some  limits in  these formulas. First consider the limit of a conducting drop: as  $Q \rightarrow \infty$ or 
$\chi \rightarrow \infty$, we have $\phi_1 (\bar{r}) = 0$ and $\phi_2 (\bar{r}) = (r^{-2}-r) \cos \theta$. Inside the drop, both
$\phi_{1\bar{r}}$ and $ \phi_{1 \bar{\theta}}$ tend to zero, that is,   the electric field is zero  for $\bar{r}<1$, but on the drop surface there is a nonzero normal component given by
$Q \phi_{1\bar{r}} \mid_{\bar{r}=1}= \phi_{2 \bar{r}}  \mid_{\bar{r}=1} =-3 \cos \bar{\theta }$.  The surface deformation function satisfies 
$\lim_{\chi \rightarrow \infty} h(\chi, Q) = (12 Q)^{-1} (9(Q-1)+9Q^{-1})$, and 
$\lim_{Q \rightarrow \infty} h(\chi, Q) = 3/4$.  In the limit $\chi \rightarrow 0$ of a dielectric drop, $\phi_1 (\bar{r}) = -(3 r/ (Q+2)) \cos \theta$  and $\phi_2 (\bar{r}) = [(Q-1)/(Q+2)] (1/r^2-r) \cos \theta$, from which the electric field is easily obtained by differentiation. The surface deformation function satisfies 
$\lim_{\chi \rightarrow 0} h(\chi, Q) = (3/4) [(Q-1)/(Q+2)]^2$.

\section{Slender body analysis}\label{slendapp}
We carry out slender-body analysis on the boundary-integral equations, following~\citet{SLB99},. Define a slenderness parameter $\epsilon = b/l$, where $l$ and $b$ are the half length and half width of drop, respectively. The existing dimensionless equations are adapted for the  slender-body scales by making the substitution
\begin{align}
S\rightarrow \frac{\epsilon}{\nu}\hat{S},\quad z\rightarrow \frac{1}{\nu}\hat{z},
\end{align}
where $\nu = R/l$ and variables with a hat are $O(1)$.

For a slender drop, i.e., $\epsilon = b/l\ll 1$, the electric field inside the drop is to leading order  in the axial direction, i.e., $E_t\approx E(z)=\nu \hat{E}(\hat{z})$. As a result, the normal stress balance (\ref{Tn})  simplifies to
\begin{align}
\frac{E_b(Q-1)}{2}(QE^2_{1n}+E^2_t)+\triangle p &= \frac{E_b(Q-1)\nu^2}{2\epsilon^2}(Q\hat{E}^2_{1n}+\epsilon^2 \hat{E}^2)+\triangle \hat{p}\nonumber\\
&\approx \frac{\nu}{\epsilon \hat{S}} - \epsilon\nu\hat{S}_{\hat{z}\hat{z}}- \frac{\alpha}{2}\phi^2_1,  \label{nsbslender}
\end{align}
where $\hat{r}= \hat{S}(z)$ is the drop surface shape and $\triangle \hat{p}$ is the constant pressure difference between drop interior and exterior.  The second term on the right hand side of (\ref{nsbslender}) is the contribution to surface tension  from the axial curvature, which is commonly retained despite being higher order.  Since $\nabla\cdot E=-\chi^2\phi_1$, the  internal field can be estimated to leading order as 
\begin{align}
\hat{E}_{\hat{r}} \approx -\frac{1}{2}\epsilon^2 \hat{S} \left(\hat{E}_{\hat{z}} + \nu^{-2}\chi^2\phi_1\right),
\end{align}
hence (see also~\cite{SLB99}),
\begin{align}
\hat{E}_{1n} \approx -\frac{\epsilon^2}{2\hat{S}} \left(\left(\hat{S}^2\hat{E}\right)_{\hat{z}} + \nu^{-2}\chi^2\hat{S}^2\phi_s\right).\label{e1n}
\end{align}


The potential exterior to the slender drop is approximated following~\citep{Acrivos78jfm}. Starting from the boundary integral representation of the exterior potential, we subtract (\ref{pe2}) from (\ref{pe1}) to obtain
\begin{align}
\phi_1(\bm{x}_0)-\phi_{\infty}+\int_{S}\phi_1\left(\frac{\p G^{\chi}}{\p n}-\frac{\p G^0}{\p n}\right) dS &= \int_{S}\frac{\p\phi_{1}}{\p n}\left(G^{\chi}-QG^0\right) dS,\label{pbie}
\end{align}
which is equivalent to the integral-equation used in~\cite{SLB99,LacHomsy07} when $\chi=0$. 
Following~\cite{Hinch:1991}, we focus on the contribution of the integral from $\epsilon \hat{S}\ll |\hat{z}-\hat{z}_0|\ll 1$. For a point on the drop centerline $(0,z_0)$, the Green's function (\ref{axhmg}) is expanded before evaluating along drop surface $\hat{r}=\hat{S}$, 
\begin{align}
G^{\chi}&=\nu\frac{e^{\left(-\frac{\chi}{\nu}\left[(\hat{z}-\hat{z}_0)^2+\epsilon^2\hat{r}^2\right]^{1/2} \right)}}{2\left((\hat{z}-\hat{z}_0)^2+\epsilon^2\hat{r}^2\right)^{1/2}}= \frac{\nu-\chi\left[(\hat{z}-\hat{z}_0)^2+\epsilon^2\hat{r}^2\right]^{1/2}+\frac{\chi^2}{2\nu}\left[(\hat{z}-\hat{z}_0)^2+\epsilon^2\hat{r}^2\right]}{2\left((\hat{z}-\hat{z}_0)^2+\epsilon^2\hat{r}^2 \right)^{1/2}}+\cdots
\end{align}
Substituting into (\ref{pbie}) yields
\begin{align}
\phi_1(\hat{z}_0) -\phi_{\infty} &+ \frac{\chi^2}{4\nu^2}\int_{-1}^1 \phi_1\frac{\epsilon^2 \hat{S}^2}{\left((\hat{z}-\hat{z}_0)^2+\epsilon^2\hat{S}^2 \right)^{1/2}}d\hat{z} \nonumber\\
= &-(1-Q)\int_{-1}^1 \frac{\hat{S} \hat{E}_{1n}}{2\left((\hat{z}-\hat{z}_0)^2+\epsilon^2\hat{S}^2 \right)^{1/2}}d\hat{z} - \frac{\chi}{2\nu} \int_{-1}^1 \hat{E}_{1n}\hat{S}d\hat{z} +\cdots \label{pbie2}
\end{align}
which is further evaluated to be
\begin{align}
\phi_1 -\phi_{\infty} &+ \frac{\epsilon^2\chi^2\ln(1/\epsilon)}{2\nu^2}\phi_1\hat{S}^2 \nonumber\\
=& (1-Q)\frac{\epsilon^2\ln(1/\epsilon)}{2}\left((\hat{S}^2\hat{E})_{\hat{z}}+\nu^{-2}\chi^2\hat{S}^2\phi_1\right) - \frac{\chi}{2\nu} \int_{-1}^1 \hat{E}_{1n}\hat{S}d\hat{z}+\cdots.\label{pbie3}
\end{align}
This is coupled with the equation for drop volume, 
\begin{align}
\int_{-1}^1 \hat{S}^2d\hat{z} = \frac{4\nu^3}{3\epsilon^2},\label{slendvol}
\end{align}
which readily yields $\nu=\epsilon^{2/3}$.

\subsection{Electric field inside a spheroid}
For a spheroid, $\hat{S}^2+\hat{z}^2 = 1$ and equation (\ref{pbie3}) with $\chi=0$ is satisfied by a uniform electric field $E_{1D}$,
\begin{align}
E_{1D} = \frac{1}{1 + \epsilon^2\ln(1/\epsilon)\left(Q-1\right)}\sim 1 - \epsilon^2\ln(1/\epsilon)\left(Q-1\right) +\cdots
\end{align}
which agrees with the approximation in~\citet{SLB99}. For order one $\chi>0$ and $\chi/\nu\gg 1$, (\ref{pbie3}) at leading order (after taking one derivative with respect to $\hat{z}$) becomes
\begin{align}
-\hat{E}_{1D}+\frac{1}{\nu} \approx -Q\frac{\epsilon^2\chi^2\ln(1/\epsilon)}{2\nu^2}\left(\phi_1\hat{S}^2\right)_{\hat{z}}.
\end{align}
Assume $\hat{E}_{1D}\sim \nu^{-1} + \Psi(\hat{z})$ so that $\phi_1 \sim -\nu^{-1}\hat{z} - \int^{\hat{z}}\Psi(s)ds$. After denoting $F=\int^{\hat{z}}\Psi(s)ds$ we arrive at
\begin{align}
-F =  Q\frac{\epsilon^2\chi^2\ln(1/\epsilon)}{2\nu^2}\left[(1-\hat{z}^2)\left(\nu^{-1}\hat{z}+F\right)\right].
\end{align}
This is rewritten as
\begin{align}
F = - K\frac{\hat{z}}{\nu}\frac{1-\hat{z}^2}{1+K(1-\hat{z}^2)}\sim- K\frac{\hat{z}}{\nu}(1-\hat{z}^2)
\end{align}
where $K=Q\frac{\chi^2\epsilon^2 \ln(1/\epsilon)}{2\nu^2}\ll 1$.
After some algebra, we arrive at an approximation for the electric field inside the drop,  which holds for $Q \ll \frac{\nu^2}{\chi^2 \epsilon^2 \ln (1/\epsilon)}$
\begin{align}
\hat{E}_{1D} &=\nu^{-1} + F_{\hat{z}} - \epsilon^2\ln(1/\epsilon)\left(Q-1\right)\nu^{-1} + \cdots\nonumber\\
&=\frac{1}{\nu} -Q\frac{\epsilon^{2}\chi^2\ln(1/\epsilon)}{2\nu^3}(1-3\hat{z}^2)  - \frac{\epsilon^2\ln(1/\epsilon)\left(Q-1\right)}{\nu} +\cdots
\end{align}
After using $E = \nu \hat{E}$, the electric field is recovered under the original  scaling to yield
\begin{align}
E_{1D} \approx 1 - \epsilon^2\ln(1/\epsilon)\left(Q-1\right) -Q\frac{\epsilon^{4/3}\chi^2\ln(1/\epsilon)}{2}(1-3(\epsilon^{2/3} z)^2). \label{e1d}
\end{align}
This is consistent with~\citet{SLB99} for $\chi=0$, and the term with $\chi$ provides a correction due to the presence of ions. The field is used to compare to the full boundary-integral simulation when the drop is elongated.

\subsection{Drop with conical ends}\label{conical}
For a drop with conical end, locally $\hat{S}\sim 1-\hat{z}$ and $\epsilon=\tan\theta_0$.
It is seen in normal stress balance (\ref{nsbslender}) that $E\sim (1-\hat{z})^{-1/2}$, then (\ref{pbie3}) shows the terms with $\chi$ serve as higher order corrections, and we still have the same equation from~\citet{SLB99}
\begin{align}
Q = 1 - \frac{8}{3\tan^2\theta_0\ln(\tan\theta_0)},\label{cuspBD}
\end{align}
which reflects a local balance of force contributions from the electric field and surface curvature.
Therefore, to leading order, the formation of a conical drop is independent of $\chi$, i.e. the influence of ions. Thus, as for a drop without electrolyte,  a conical end is only expected when $Q$ is sufficiently large regardless of $\chi$. To be specific,  ~\citet{SLB99} gives a minimum $Q$ around $15.5$, above which conical tip is possible. 

Finally, we note that the slender drop shape can be analyzed by coupling (\ref{pbie3}), (\ref{nsbslender}) and (\ref{slendvol}) (see \citet{Sherwood91,SLB99,RY2010}). Our preliminary results show a mild singularity between a conical and a rounded end exists (same in  \cite{Sherwood91,RY2010}) and $E_b(Q-1)\epsilon^{7/3}\ln(1/\epsilon)\sim O(1)$. However, we do not pursue this further in current study as the predicted shape using slender-body is usually in poor comparison with the full simulation.




\end{document}